\documentclass[fleqn,usenatbib]{mnras}

\usepackage[T1]{fontenc}
\usepackage{ae,aecompl}
\usepackage{flushend}

\usepackage{graphicx}	
\usepackage{amsmath}	
\usepackage{amssymb}	
\usepackage{bm,xcolor}
\usepackage{lmodern,slantsc}
\usepackage{booktabs}
\usepackage[normalem]{ulem} 
\usepackage{comment}

\newcommand{\de}{\mathrm{d}}

\newcommand{\hi}{H\textsc{i}}
\newcommand{\secref}[1]{\hyperref[#1]{Section~\ref*{#1}}}
\newcommand{\appref}[1]{\hyperref[#1]{Appendix~\ref*{#1}}}

\title[BAO: \hi\ IM BAO cross-correlations in multipoles]{Baryon acoustic oscillations from H{\Large I} intensity mapping: the importance of cross-correlations in the monopole and quadrupole}
\author[Rubiola, Cunnington, \& Camera]{Andrea Rubiola,$^1$\thanks{\texttt{andrea.rubiola97@gmail.com}}
Steven Cunnington,$^{2,3}$\thanks{ \texttt{steven.cunnington@ed.ac.uk}}
and Stefano Camera$^{1,4,5}$\thanks{\texttt{stefano.camera@unito.it}}
\\
$^{1}$Dipartimento di Fisica, Universit\`a degli Studi di Torino, via P.\ Giuria 1, 10125 Torino, Italy\\
$^{2}$Institute for Astronomy, The University of Edinburgh, Royal Observatory, Edinburgh EH9 3HJ, UK\\
$^{3}$School of Physics \& Astronomy, Queen Mary University of London, 327 Mile End Road, London E1 4NS, UK\\
$^{4}$INFN -- Istituto Nazionale di Fisica Nucleare, Sezione di Torino, via P.\ Giuria 1, 10125 Torino, Italy\\
$^{5}$Department of Physics \& Astronomy, University of the Western Cape, Cape Town 7535, South Africa
}

\date{Accepted XXX. Received YYY; in original form ZZZ}

\pubyear{2021}

\begin{document}
\label{firstpage}
\pagerange{\pageref{firstpage}--\pageref{lastpage}}
\maketitle
\begin{abstract}
Cosmological parameter estimation in the post-reionisation era via neutral hydrogen radio emission ($\text{\hi}$), is among the key science goals of the forthcoming SKA Observatory (SKAO). This paper explores detection capability for baryon acoustic oscillations (BAO) with a suite of \(100\) simulations introducing the main limitations from foreground contamination and poor angular resolution caused by the radio telescope beam. Such broad single-dish beam representing a serious challenge for BAO detection with \hi\ intensity mapping, we investigate a multipole expansion approach as a means for mitigating such limitations. We also showcase the gains made from cross-correlating the \hi\ intensity mapping data with an overlapping spectroscopic galaxy survey, aiming to test potential synergies between the SKA Project and other future cosmological experiments at optical/near-infrared wavelengths. For our ${\sim}\,4\,000\,\mathrm{deg}^2$ data set at $z\,{=}\,0.9$, replicating the essential features of an SKAO \hi\ intensity mapping survey, we were able to achieve a ${\sim}\,4.5\sigma$ detection of BAO features in auto-correlation despite the dominant beam effect. Cross-correlation with an overlapping galaxy survey can increase this to a ${\sim}\,6\sigma$ detection. Furthermore, including the power spectrum quadrupole besides the monopole in a joint fit can approximately double the BAO detection significance. Despite not implementing a radial-only $P(k_\parallel)$ analysis in favour of the three-dimensional $P(\bm{k})$ and its multipoles, we were still able to obtain robust constraints on the radial Alcock-Paczynski parameter, whereas the perpendicular parameter remains unconstrained and prior dominated due to beam effects. 
\end{abstract}


\section{Introduction}
The 2020s and 2030s will witness a golden age for cosmological experiments surveying the large-scale cosmic structure, picking up the legacy of the various missions of the past decades dedicated to the study of the cosmic microwave background radiation, which provided us with the most precise measurements of cosmological fundamental quantities so far \citep[e.g.][]{COBE_Fixsen_1996,WMAP_Hinshaw_2013,Akrami2018}. Among them, we can quote the European Space Agency's \textit{Euclid} satellite \citep{Laureijs:2011gra,Amendola2013,Amendola2016}, the Nancy Grace Roman Space Telescope \citep{2015arXiv150303757S}, the Spectro-Photometer for the History of the Universe, Epoch of Reionization, and Ices Explorer \citep[SPHEREx,][]{2014arXiv1412.4872D,2018arXiv180505489D}, the Dark Energy Spectroscopic Instrument (DESI) \citep{2016arXiv161100036D,2016arXiv161100037D}, the Vera C.\ Rubin Observatory \citep{2009arXiv0912.0201L,2018arXiv180901669T,Ivezic:2008fe}, and the SKA Observatory (SKAO) the SKA Observatory (SKAO) \citep{Maartens:2015mra,SKA_Redbook_2020}.

Due to the emission of highly energetic light coming from population-III stars, the neutral hydrogen (\hi) in the intergalactic medium sees a massive depletion during the epoch of reionisation, reducing its contribution from being a dominant component to a present-day abundance of $\Omega_{\text {HI}} \sim 10^{-4}$. One of the most pressing science cases for the SKAO's radio-telescope is then looking for the \hi\ that survived  into the subsequent post-reionisation epoch ($z<5$). This can be achieved by measuring its characteristic 21-cm transition, resorting to a long-known astrophysical observable, whose applications date back to the 1950s \citep[][]{HI_1958_first,Furlanetto:2004ha,Furlanetto_2006}.
Since \hi\ overwhelmingly resides inside galaxies in the post-reionisation epoch, it acts as a tracer of galaxy clustering and hence of the underlying large scale cosmic structure. The weakness of the signal, demanding high technological standards, has hitherto limited the use of this observable in cosmology: the SKAO will be able to observe such spectral line via so-called `intensity mapping' \citep{Bharadwaj:2000av,Battye:2004re,Wyithe:2007rq,Chang:2007xk}, a technique devised to collect large amounts of signal faster than galaxy surveys. The technique has seen a growing number of cosmological signal detections \citep{Masui:2012zc,Anderson:2017ert,Wolz:2021ofa,Cunnington:2022uzo} and the recent successful demonstration of single-dish intensity mapping calibration with an array of dishes \citep{Wang:2020lkn} lays the foundation for advancing the maturity of this technique with the SKAO and its pathfinder surveys such as MeerKAT, a 64 dish precursor to the SKAO \citep{Santos:2017qgq,Pourtsidou:2017era}.

On the much shorter wavelengths of the near-infrared and optical bands, one of the main probes of \textit{Euclid} and \textit{Roman} will consist of spectroscopy to detect the H$\alpha$ emission line in galaxies between, in particular between $z=0.9$ and $2.0$. The resulting galaxy catalogues, amounting to some tens of million objects with spectroscopic redshift estimates, will allow us to track the growth and clustering of cosmic structures following the path carved by previous collaborations \citep[e.g.][]{GilMarin2016,Zhao2017,Chuang2017,pellejeroibanez2016clustering,Wang2017}.

Both optical and radio surveys are expected to collect a large amount of data, in the order of the yearly Internet traffic \citep[][]{Farnes_2018}. The forecasting effort needed to optimise their performance is therefore, clearly, of utter importance. More specifically, our work revolves around the detection forecast of baryon acoustic oscillations (BAO), considered one of the primary goals of any future cosmological survey, to be attempted with both conventional methods, e.g.\ \textit{Euclid} \citep[][]{EuclidVII:2020}, DESI, and more novel techniques, like \hi\ intensity mapping with the SKAO \citep[][]{SKA:BAO}. Furthermore, pathfinder surveys like MeerKAT will also aim to provide the first measurements of BAO using the intensity mapping technique \citep{Santos:2017qgq}.

BAO originate from the past interaction of radiation and baryonic matter, which ended at decoupling ($z\sim1000$). Until that time, radiation pressure sustained the baryon gravity, preventing their collapse and inducing acoustic oscillations with a sound horizon $r_{\text {s}}=c_{\text {s}} t$, the sound speed being
\begin{equation}
c_{\text s}\approx 1.15 c \sqrt{\frac{\rho_{\text {rad}}}{4\rho_{\text {rad}}+3\rho_{\text {b}}}}\;, \label{eq:sound_speed} 
\end{equation}
where $\rho_{\rm rad}$ and $\rho_{\rm b}$ respectively denote the radiation and baryon energy densities. Evaluating this at the decoupling redshift we recover a sound horizon scale of $r_\text{s} \sim 105\,\mathrm{Mpc}\,h^{-1}$, which is retained in the matter clustering as a preferential scale of separation between galaxies, even after baryons realign with dark matter. As a result, we find in the 2-point correlation function a secondary probability excess (a `bump') at the aforementioned scale (see \autoref{fig:correlationfunction}), which has been observed in galaxy surveys for over two decades \citep{Percival:2001hw,Eisenstein_2005_BAO}.

Taking the Fourier transform of the correlation function provides the power spectrum in which the BAO bump translates into a series of characteristic `wiggles' in the $k=[0.02,0.3]\,h\,\mathrm{Mpc}^{-1}$ interval, whose detection  will be investigated throughout this paper.
BAO represent an interesting cosmological observable: having well defined radial and transverse dimensions, they can be used as a standard ruler analogous to standard candles, allowing estimations of the Hubble parameter and of other cosmological parameters involved in the radial and angular distance functions. Furthermore, they are a relatively large-scale phenomenon, difficult to be mimicked or deformed by other physical processes---apart from the well-known smoothing out of the wiggles due to the transition to non-linear scales \citep[][]{Crocce_BAO_2008,Sugiyama_BAO_2014}. Hence, BAO are robust and can be consistently modelled with linear theory. 

However, detecting BAO with \hi\ intensity mapping introduces a set of unique challenges \citep[see e.g.][]{Battye:2012tg}. Since \hi\ maps the diffuse, unresolved \hi\ emission, observations will also accumulate any other radiation in the same frequency range as the redshifted \hi. Perhaps most contaminating is from 21cm foregrounds which can dominate the \hi\ signal by several orders of magnitude \citep{Wolz:2013wna,Alonso:2014sna,Cunnington:2020njn}. Furthermore, the angular resolution of the maps can be limited by the beam size of the instrument, which for the case of a single-dish intensity mapping experiment such as SKAO's, can be quite broad, of order $1\,\mathrm{deg}$ in size at $z=0.4$ \citep{SKA_Redbook_2020}. These observational effects are expected to impact the statistical recovery of the BAO in intensity mapping \citep{Villaescusa-Navarro:2016kbz,Kennedy:2021srz,Avila:2021wih}. On the other hand, cross-correlations of the \hi\ intensity mapping signal with galaxy clustering appear as a viable alternative: a typical galaxy survey will not be limited by angular resolution, nor is it affected by foreground removal \citep[though see e.g.][]{2019JCAP...04..023M}. Thus these limitations to the \hi\ intensity mapping method could be mitigated in cross-correlation \citep[see studies in e.g.][]{Wolz:2015ckn, Pourtsidou:2016dzn}.

In this work we look to extend upon previous investigations of BAO detection with \hi\ intensity mapping \citep{Villaescusa-Navarro:2016kbz} by including a galaxy survey cross-correlation, in comparison with the \hi\ auto-correlation. We use simulation-based data sets inclusive of the relevant intensity mapping observational effects, namely foreground contamination, a broad single-dish telescope beam, and thermal noise; we then evaluate the BAO detection significance along with performing cosmological parameter estimation, to evaluate the merit of each approach. Unlike the previous work of \citet{Villaescusa-Navarro:2016kbz}, which employed a radial power spectrum $P(k_\parallel)$, we use the conventional three-dimensional power spectrum $P(\bm{k})$. This allows us to utilise a multipole expansion of the power spectrum  which leads to improvements in BAO detection, compared to a single-fit to the monopole. This inclusion of higher-order multipoles was investigated in recent works \citep{Kennedy:2021srz,Avila:2021wih} in configuration space. 
\citet{Avila:2021wih} 
also extended their investigation into using clustering $\mu$-wedges in chosen regions to filter out the unwanted systematic effects from 21cm foregrounds and the beam. Our approach differs by using the power spectrum in Fourier space in a simulations-based test, to demonstrate the importance of including the quadrupole in a joint fit with the monopole. 
Previous studies have investigated a Fourier space analysis \citep{Soares_Cunnington2021} but looked at fitting the full-shape power spectrum. In this work we exclusively constrain the BAO features to see if this is feasible with a low resolution \hi\ intensity mapping experiment. Detection of cosmological features will generally benefit the maiden demonstration of \hi\ IM in auto-correlation, as yet to be achieved. A confident detection is difficult when fitting for a featureless power spectrum where contribution to the amplitude comes from both \hi\ signal and additive biases from non-cosmological residuals and systematics, as discussed in \citet{Cunnington:2022ryj}. Furthermore, we also forecast the added benefit from a cross-correlation with an overlapping galaxy survey.

The paper is structured as follows: in \secref{sec:sim_setup} we describe the criteria our simulation setups are based on and the most important features of our fitting models.
Our data analysis results are collected in \secref{ref:results_sec}; finally, we wrap up our results and their discussion in \secref{sec:disc}. For the sake of better readability, most of the standard mathematical formalism adopted in our work is collected in  \appref{Append_ref_pk} and \ref{Appendix_uncert}, together with the study of the signal-to-noise ratio (SNR) and uncertainties in  \appref{sec:uncert_snr_section}. Conversely, less frequent concepts or formulas whose exposition is necessary for a better understanding of the topic will be discussed in their proper context.

\section{Methodology}\label{sec:sim_setup}
In this section, we describe the methodological approach we followed to generate our mock data sets, inclusive of observational effects such as foreground contamination, beam smoothing, and instrumental noise, as well as the statistical analysis to analyse it.

\subsection{Cosmological Simulations}
We begin by producing log-normal simulations onto a Cartesian grid, based on a methodology first introduced by \citet{Coles_Jones_Lognormal}, following the prescription outlined in \citet{Beutler_lognormal}. In this process, we utilise the publicly available codes of \texttt{Nbodykit}\footnote{\href{https://nbodykit.readthedocs.io/en/latest/index.html}{nbodykit.readthedocs.io}} \citep{Nbody_Hand} and \texttt{hmf}\footnote{\href{https://github.com/halomod/hmf}{github.com/halomod/hmf}} \citep{hmf}, which both integrate the Boltzmann solvers \texttt{CLASS} \citep[][]{lesgourgues2011_Class,Blas_2011_class}  and \texttt{CAMB} \citep[][]{Lewis:1999_CAMB} 
respectively. We set up our log-normal simulations at the three redshift values of $z=[0.9, 1.35, 2.0]$. This choice is motivated by forthcoming intensity mapping and optical galaxy surveys, with $z=0.9$ and $z=2.0$ bracketing the redshift coverage of a \textit{Euclid}-like spectroscopic survey, and $z=1.35$ being an intermediate value based on a MeerKAT (SKAO's precursor) UHF band survey, which ends at $z=1.45$.

The simulated data are designed not to have excessive depth along the $\mathrm{z}$-axis.\footnote{Note that we orientate our Cartesian grid so that the $\mathrm{z}$-axis points along the line-of-sight direction. Moreover, to avoid confusion between the Cartesian $\mathrm{z}$-axis and redshift $z$, we adopt Roman letters to denote Cartesian coordinates, namely $(\mathrm{x},\,\mathrm{y},\,\mathrm{z})$.} 
The percentage variation of the relevant quantities (fitting parameters, galaxy densities and noise PS intensities) in the mock redshift depth compared to the value at the nominal redshift are presented in \autoref{tab:parameters_nominal_2}.
This is to ensure our assumption that parameters will not vary within a particular redshift bin is reasonably valid. 
At the same time, the simulation box cannot be arbitrarily small due to physical reasons: we must properly sample the BAO scales around $k\sim0.02\,h\,\mathrm{Mpc}^{-1}$. 
A good compromise for all redshifts is a depth along the $z$-axis of $l_\mathrm{z}=600\,\mathrm{Mpc}\,h^{-1}$, which we use in each redshift case. To ensure a good level of signal-to-noise ratio (SNR) given the limited depth, we set our angular dimensions along the $\mathrm{x}$- and $\mathrm{y}$-axis to a length of $l_\mathrm{x}=l_\mathrm{y}=2400\,\mathrm{Mpc}\,h^{-1}$. We use the same physical size in each redshift bin to ensure a like-for-like comparison. These angular sizes respectively correspond to $4180,\,2352$, and $1460\,\mathrm{deg}^2$ of observed sky or, equivalently, to sky fractions $f_\text{sky}=0.12,\,0.07,\,0.05$. All these values are within the targeted sky sizes of e.g.\ MeerKLASS \citep{Santos:2017qgq}, \textit{Euclid}, and \textit{Roman}. We find this field of view sufficient to ensure a detection.

Our simulations are then outlined according to the following steps:
\begin{enumerate}
    \item We divide each field into a uniform grid of $N_{\text {grid}}=\{400,\,400,\,400\}$ voxels, which allows all scales of interest to be sufficiently sampled.
    \item The power spectrum underlying the simulation is chosen to be non-linear for the sake of greater realism and to encapsulate some of the limitations to the BAO features caused by non-linear effects, without resorting to expensive $N$-body simulations. To achieve this we use \texttt{Nbodykit} to calculate two linear power spectra, one with BAO and the other with a `no-wiggles' (also, broad-band) power spectrum based on the transfer functions in \citet{Eisenstein:1997ik}. The power spectrum with BAO is then used for the simulation. The role of the broad-band counterpart will be described later in the paper. We then use \texttt{hmf} to convert both power spectra to the non-linear version with the standard prescription \texttt{halofit} \citep{Halofit_2003_original,Halofit_Takahashi_2012}.    
    \item We introduce additional features to the input power spectra, such as redshift-space distortions (RSD), modelled according to \citet[][see \appref{Append_ref_pk} for more explicit details and discussion on this formalism]{Kaiser:1987qv}. We then use the input power spectra to produce log-normal density fields with our mock generator. Specifically, we employ 100 mocks at all three redshifts, 
    following the method suggested by \citep[][]{Villaescusa-Navarro:2016kbz}  
    and subsequently checked to provide statistically sound results.
    \item For the \hi\ intensity maps, we add a \hi\ linear bias to the field and add astrophysical foregrounds (see \secref{sec:FG+noisesec}). The field is then smoothed in the transverse direction, to imitate SKAO beam effects, then we apply a thermal noise to the map. Lastly, the $\delta_{\mathrm{\hi}}$ field is foreground cleaned via principal components analysis (PCA). This should emulate some residual foreground contamination and signal-loss from the foreground clean expected in a real survey experiment.
    \item For the galaxy maps, we Poisson sample each generated matter field with a galaxy number density calculated according to the model outlined in \citet[][]{EuclidVII:2020}, finding, for the three chosen redshifts, the following $\bar n_\mathrm{gal} = \{7.2,\,4.6,\,1.1\}\times 10^{-4}\, h^3\,\mathrm{Mpc}^{-3}$ values. 
    \item 
    The simulations should take into account the noise contributions. The sampling shot noise is not relevant for the \hi\ auto-correlation, whilst the galaxy count shot noise should be almost entirely suppressed in the cross-correlation. (See \autoref{tab:parameters_nominal_2}, deriving from the galaxy density model necessary to initialise the simulations and presented in \appref{sec:shot_noise}.) Similarly, the thermal noise levels (also quoted in \autoref{tab:parameters_nominal_2} and discussed in \appref{sec:thermal_noise}) follow from standard models \citep{SKA_Redbook_2020}. We work under the hypothesis of $10\,000$ single-dish observation hours and prove to be well below the power spectrum signal at the scales of interest. It is also highly unlikely thermal noise levels will be an issue with the SKAO, and it will instead be the large beam which is the dominant challenge for BAO detection.
    \item Finally, we apply a power spectrum estimator pipeline 
    see \appref{Append_ref_pk})
    to all the generated data sets, both for \hi\ auto-correlation and \hi\,galaxy cross-correlation. Since we intend to explore the benefit from a multipole expansion formalism in this work, we also measure the quadrupole of each data set. Definitions for these can be found in \autoref{eq:ell_autoHI} and \autoref{eq:ell_cross}.
    During this latter stage, we also extract from the 100 samples the data variances and covariances that allow us to estimate the empirical uncertainties. They are compared with the theoretical values in \appref{Appendix_uncert}: the excellent agreement we find ensures we can assign the theoretical values as error bars to our data. Notice that, although the signal is automatically free from the noise in our setup, we need a proper model for $P_{\rm{th}}$ and $P_{\rm{shot}}$  because they appear in the uncertainty calculation.
\end{enumerate}
\begin{table}
\centering
\caption{Percentage variation of fitting and simulation parameters with respect to nominal values in each redshift bin. The fixed bin width of $\sim600\,h^{-1}\,\mathrm{Mpc}$ assigned to every simulation respectively corresponds to $\Delta z=0.115$, $0.148$, and $0.205$.}
\begin{tabular}{l c c c }
\hline
Parameter & \multicolumn{3}{c}{$\Delta\,[\%]$} \\
\cline{2-4}
& $z=0.9$ & $z=1.35$ & $z=2.0$\\
\hline\hline
$R\,[h^{-1}\,\mathrm{Mpc}]$ & $37.2$ & $28.8$ & $28.6$ \\
$\sigma_r\,[h^{-1}\,\mathrm{Mpc}]$ & $-1.5$ & $-3.2$ & $-5.1$ \\
$n_{\rm gal}\,[(h^{-1}\,\mathrm{Mpc})^{-3}]$ & $-6.1$ & $-35.5$ & $-67.4$ \\
$P_{\rm shot}\,[(h^{-1}\,\mathrm{Mpc})^3]$ & $6.5$ & $54.9$ & $207.1$ \\
$b_{\rm gal}$ & $14.9$ & $11.4$ & $11.2$ \\
$b_{\rm\hi}$ & $6.5$ & $8.0$ & $12.0$ \\
$b_{\rm\hi}\overline{T}_{\rm{b}}\,[\mathrm{mK}]$ & $27.3$ & $26.2$ & $32.3$ \\
$P_{\rm th}\,[\mathrm{mK}^2\,(h^{-1}\,\mathrm{Mpc})^3]$ & $11.1$ & $8.9$ & $8.8$\\
\hline
\end{tabular}
\label{tab:parameters_nominal_2}
\end{table}

\subsection{Fitting template and parameter space}
We will make use of three different fitting strategies, in every case resorting to the the Markov chain Monte Carlo (MCMC) sampler \texttt{emcee} \citep{emcee}. Specifically, we shall fit the monopole alone ($P_{0}$), the quadrupole alone ($P_{2}$), and the monopole and quadrupole jointly, ($\left[P_{0},P_{2}\right]$ or `joint fit'); the three cases can all be formally written as
\begin{equation}
    -2\ln\mathcal{L}=(\bm{P}_\text{data} - \bm{P}_\text{model})^{\sf T}\mathbfss C^{-1}( \bm{P}_\text{data} - \bm{P}_\text{model})\;,\label{eq:likelihood_general}
\end{equation} 
$\mathbfss C^{-1}$ being the inverse of the data covariance matrix. What changes between the first two cases and the joint fit is that in the former the posterior probabilities are calculated by considering the sole variance of the multipole with respect to itself, whereas the latter case exploits the combination of the monopole and quadrupole variance terms plus their covariance. Such a simultaneous fit should tighten and improve the posterior distributions.

We make our fitting template dependent upon the pair of variables ($\mu$, $k$), with $\mu$ the cosine of the angle between the line-of-sight direction and the wave-vector ${\bm k}$, and $k^2=k_{\parallel}^2+k_{\perp}^2$  the modulus of each mode in the Fourier space. We use the Alcock-Paczynski (AP) formalism \citep{Alcock:1979mp} 
based on the relations
\begin{align}
\alpha_{\parallel}=\frac{H^{\text{fid}}(z)}{H(z)}\;,\label{eq:AP_fact_par}\\
\alpha_{\perp}=\frac{D_{\text{A}}(z)}{D_{\text{A}}^{\text{fid}}(z)}\;,
\label{eq:AP_fact_perp}
\end{align}
where the $\text{fid}$ superscript denotes the fiducial values calculated with the reference cosmology for the underlying power spectrum, in our case a vanilla $\Lambda$CDM model \citep[][]{Ade:2015xua}. Here, $H(z)$ and $D_{\rm A}(z)$ denote respectively the Hubble rate and the angular distance at redshift $z$.

Both AP parameters are expected to be $\approx 1$ when the cosmology underlying the fitting template and data agree.
Thus, we can redefine the transverse and the radial fiducial modes, $k_{\perp}^{\text{fid}}$ and $k_{\parallel}^{\text{fid}}$, as $k_{\perp}=k_{\perp}^{\text{fid}}/\alpha_{\perp}$ and $k_{\parallel}=k_{\parallel}^{\text{fid}}/\alpha_{\parallel}$.
Recalling that $\mu=k_{\parallel}/k$, the independent variables of the power spectrum can be combined in order to write \begin{equation}
k=\frac{k^{\text{fid}}}{\alpha_{\perp}}\left[1+(\mu^{\text{fid}})^2(F_{\rm{AP}}^{-2}-1)\right]^{1/2}\;, \label{k_f_to_k}
\end{equation} 
which depends on $k^{\text{fid}}$, on the ratio $F_{\textrm{AP}}=\alpha_{\parallel}/\alpha_{\perp}$, and on $\mu^{\text{fid}}$; the cosine $\mu$ can then be made dependent on $\mu^{\text{fid}}$ according to
\begin{equation}
\mu=\frac{\mu^{\text{fid}}}{F_{\rm{AP}}}\left[1+(\mu^{\text{fid}})^2(F_{\rm{AP}}^{-2}-1)\right]^{-1/2}\;.
\label{mu_f_to_mu}    
\end{equation}
Hence, the final fitting template for the power spectrum reads
\begin{multline}
P_{\text {fit}}(k,\mu,z)=P_{\text {nw}}(k,\mu,z)+\frac{1}{\alpha_{\perp}^2\alpha_{\parallel}}\\
\times \left\lbrace P_{\text w}\left[k\left(\alpha_{\parallel},\alpha_{\perp}\right),\mu,z\right]-P_{\text {nw}}\left[k\left(\alpha_{ \parallel},\alpha_{\perp}\right),\mu,z\right]\right\rbrace\;,\label{fit_template}
\end{multline}
where, depending on whether we are looking at auto- or cross-correlations, the input models are provided in \autoref{eq:ell_autoHI} and \ref{eq:ell_cross}. The non-linear matter power spectrum either contains BAO for $P_{\text w}$, or is the broad-band (`no-wiggles') power spectrum version for $P_\text{nw}$. We look for the first  three wiggles of the BAO signal (the most visible ones in the power spectrum) in the interval $k=[0.02,0.2]\,h\,\mathrm{Mpc}^{-1}$. We also fit a pure `no-wiggles' counterpart, in order to establish BAO detection significance.

The parameter sets associated to \autoref{fit_template} are \begin{align}
    \Theta_{\text  {auto}}&=\{\alpha_{\parallel},\alpha_{\perp},b_{\text  {\hi}}\overline{T}_{\text  b},R,n_{\parallel},n_{\perp}\}\;,\\
    \Theta_{\text {cross}}&=\{\alpha_{\parallel},\alpha_{\perp},b_{\text  g},b_{\text  {\hi}}\overline{T}_{\text b},R,\sigma_{\text  r},n_{\parallel},n_{\perp} \}\;.
\end{align}
They correspond to: the AP factors; the (effective) bias of the various tracers; the radio-telescope transverse resolutions (and radial resolution for spectroscopic measurements); and the foreground-cleaning compensation factors. Their meaning and underlying models are again described in greater detail in \appref{Append_ref_pk}, whilst the fiducial values at each mock nominal redshift are tabulated in \autoref{tab:parameters_nominal_2}.
Finally, note that the pure `no-wiggles' model contemplates the same parameter space, except for $\alpha_{\parallel}$ and $\alpha_{\perp}$.

As a final remark, whilst the MCMC alone permits an evaluation of the goodness-of-fit, based on the comparison between priors and posteriors, we also calculate $\chi^2=-2\ln\mathcal L$ and its value normalised to the number of degrees of freedom, $\chi^2/\text{dof}$, assuming as best-fit values the samples at the 50th percentile of the posterior distributions. This provides a useful framework to assess the level of significance for the BAO detection.

\section{Results} \label{ref:results_sec}
\begin{figure*}
    \centering
    \includegraphics[width=2.1\columnwidth]{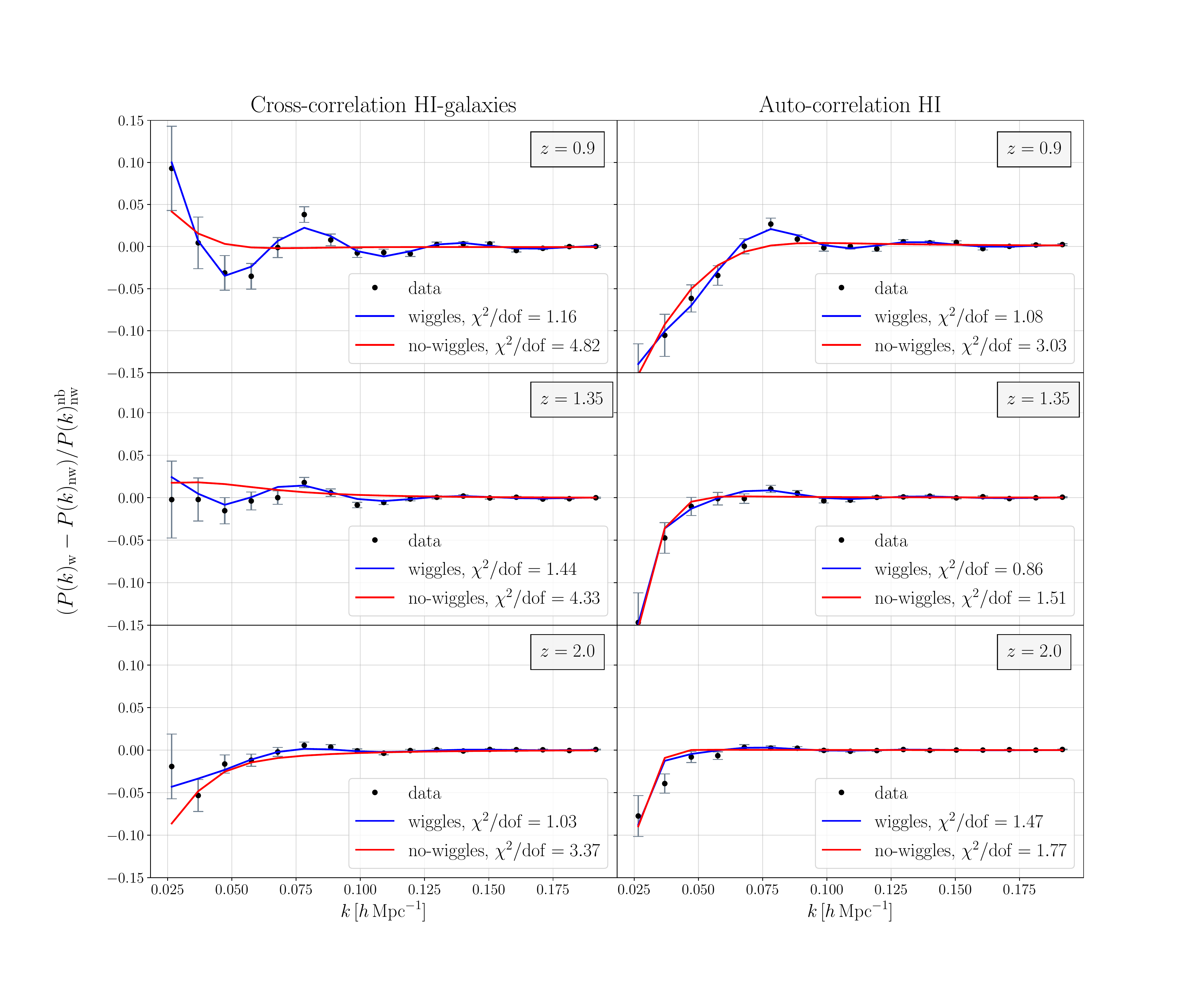}
    \caption{A single chosen realisation at each redshift of the cross-correlation (left-column) and auto-correlation (right column) monopole power spectra, at each redshift. Overlaid are the models including BAO wiggles (blue-line) and without (red-line) along with their corresponding $\chi^2/\text{dof}$ fit to the data, displayed in each legend.}
    \label{fig:stacked_fits_mono}
\end{figure*}

In this section, we will show some best-fit plots and explore the data analysis results. 
As a general convention, the power spectrum fits displayed in \autoref{fig:stacked_fits_mono} and \ref{fig:stacked_fits_mono_quadru} (cross-correlation on the left, auto-correlation on the right) are normalised for each tracer
by subtracting the fit for the no-wiggles $P_{\rm nw}(k)$ and by dividing the result by the no-beam (hence the $\text{nb}$ superscript), no-wiggles power spectrum $P_{\rm nw}^\text{nb}(k)$. This way, we can demonstrate how much the BAO exceed the broad-band power spectrum at every redshift, taking into account the smoothing effects. Concerning the error bars, the discussion on their evaluation is presented in \appref{Appendix_uncert}.

Where we show the posterior distribution and marginalised contours for both cross- and auto-correlation, we only show the common parameters, that is $b_{\text g}$ and $\sigma_r$ are not displayed. In such plots, the dashed vertical line corresponds to the fiducial values of the parameters.

\subsection{Monopole}
Some qualitative observations can already be made from \autoref{fig:stacked_fits_mono}, where we observe the progressively reducing amplitude of the wiggles.
The first `bump', however small, remains recognisable by eye, whilst the following ones are progressively less distinguishable from the broad-band counterpart, consistently with the $k$-dependence of the beam. At the same time, it seems that the cross-correlation allows for a slightly better resolution of the secondary oscillations with respect to the auto-correlation, due to the absence of transverse smoothing in the galaxy sample, which mitigates the impact from the beam. 

\begin{figure}
    \begin{center}
    \includegraphics[width=1\columnwidth]{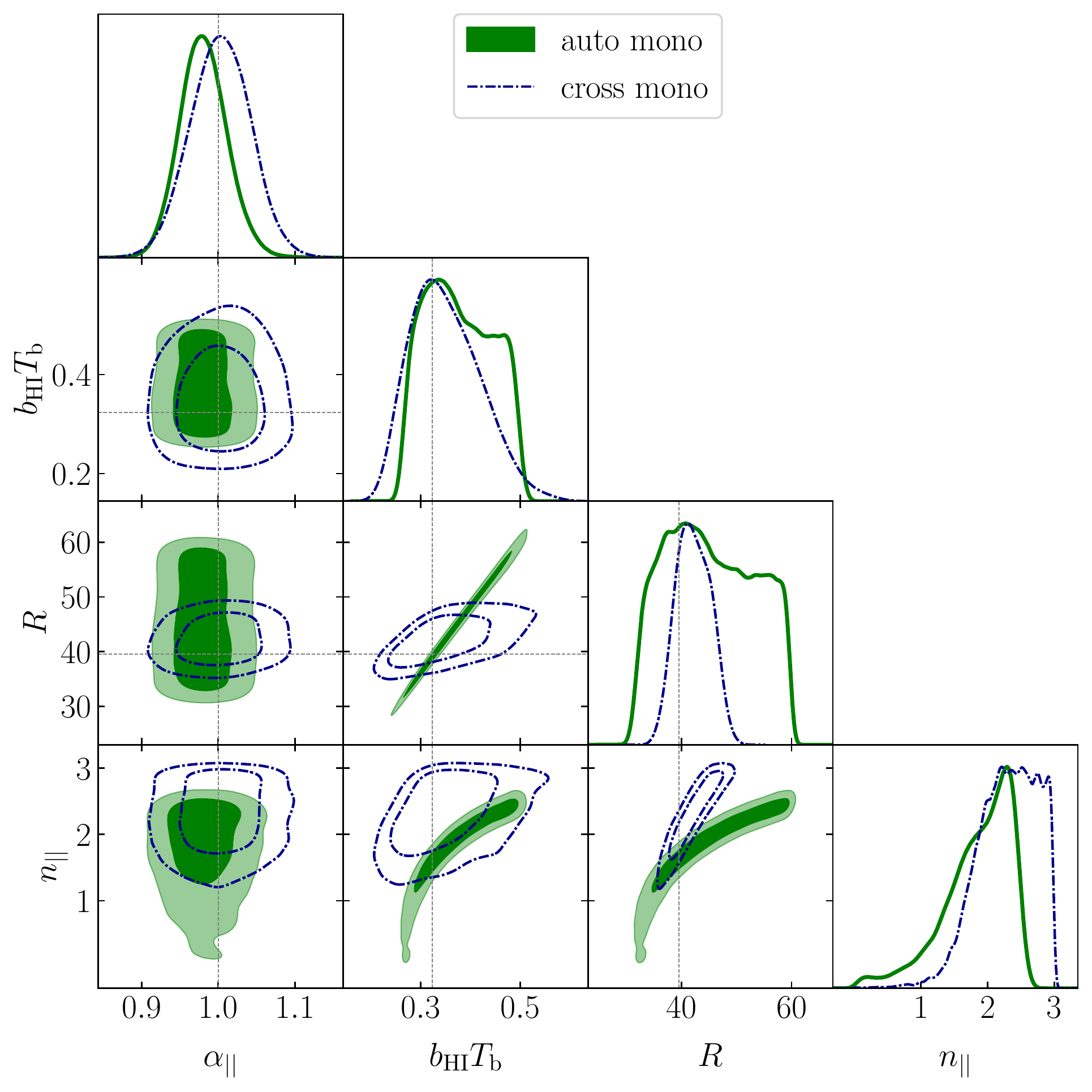}
    \end{center}
    \caption{Example monopole posteriors for the shared parameters for a single chosen realisation in auto- and cross-correlation at $z=1.35$.}
    \label{fig:cross_auto_corner135_fit}
\end{figure}

\begin{figure}
    \includegraphics[width=\columnwidth]{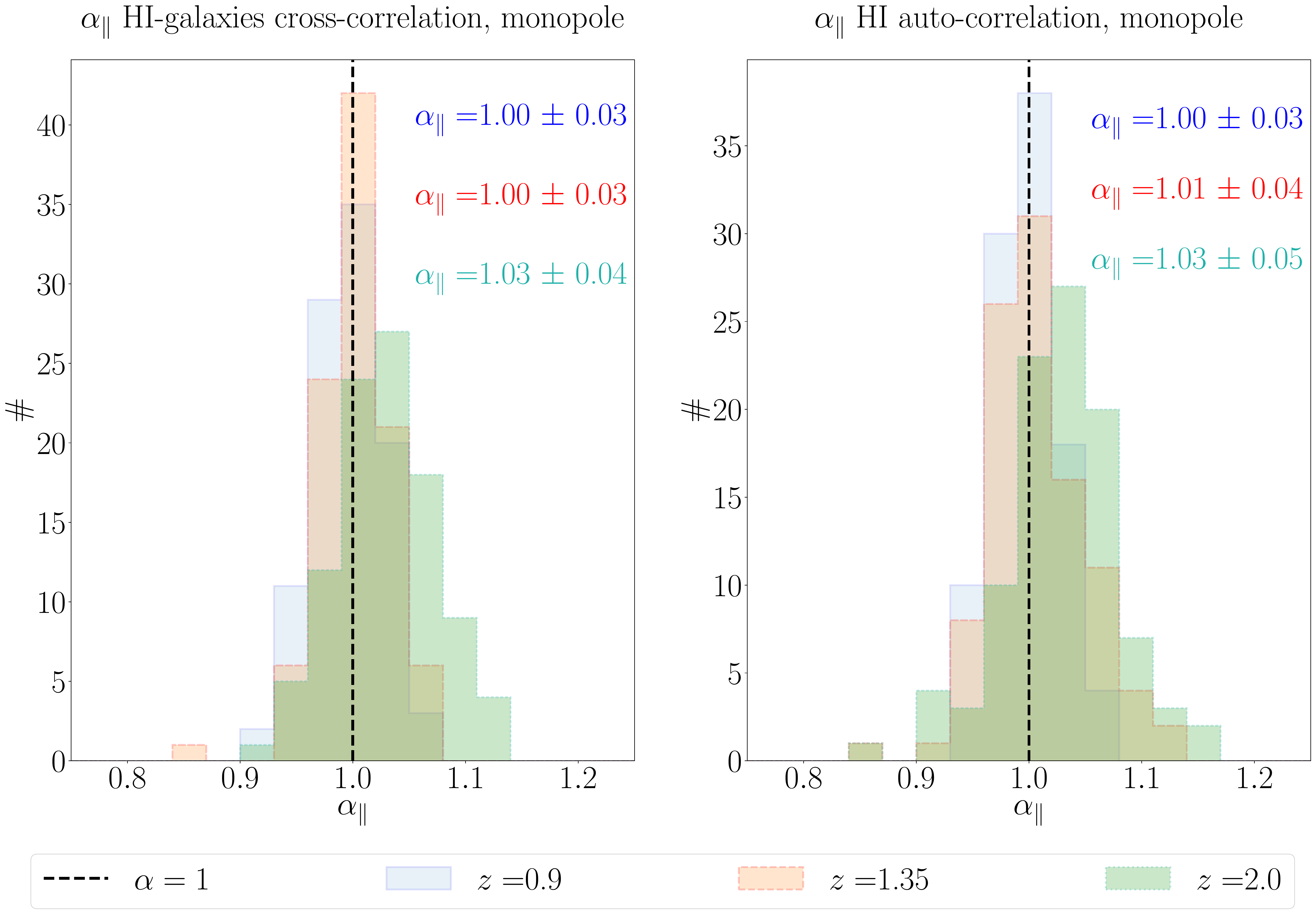}
    \caption{Distribution of the best fit $\alpha_\parallel$
    values for all realisations at different redshifts for both auto- and cross-correlation monopole.}
    \label{fig:alpha_mono}
\end{figure}

We show the monopole posteriors for both auto- and cross-correlation in \autoref{fig:cross_auto_corner135_fit}. We observe a good level of independence between $\alpha_\parallel$ and the rest of the parameter space, which we can interpret as the possibility of retrieving robust cosmological information decoupled from nuisance and astrophysical variables.
Due to its direct link with the goal of testing the cosmological model, in \autoref{fig:alpha_mono} we include the distribution of best-fit $\alpha_{\parallel}$ parameters from our suite of simulations for both methods and at all redshifts. We see good $\alpha_\parallel$ constraints in both auto- and cross-correlation, with a slight broadening of the distribution for higher redshift, shown by the quoted $\pm$ $68\%$ confidence intervals from the mean in the top-right of each panel. Most likely due to the beam limited transverse resolution, no significant information on the $\alpha_{\perp}$ parameter can be retrieved, other than a uniform distribution corresponding to the chosen prior interval. The use of the 50th percentile value as a best-fit parameter does not imply in this case any strong constraining of the measurement. This is also why we have omitted the $\alpha_\perp$ posterior in \autoref{fig:cross_auto_corner135_fit}, which showed highly prior dominated results, indicating that the perpendicular scales are severely hampered by the radio telescope beam. This unfortunately shows little improvement in cross-correlation.

Concerning the inability to constrain the perpendicular $\alpha_\perp$ parameter, it is interesting to compare this with the conclusions from \citet{Soares_Cunnington2021}. Their work also looked at constraining the AP parameters with \hi\ intensity mapping, but instead used a model fit to the full-shape power spectrum. Unlike this work, they found reasonable constraints could be achieved on $\alpha_\perp$, despite the presence of a broad intensity mapping beam, albeit a smaller size than what we consider in this work. Alongside the difference in beam size, the main reason for our work not achieving any constraints on $\alpha_\perp$ will be down to our method of modelling the wiggles, relative to the no-wiggles power spectrum. As we have seen from \autoref{fig:stacked_fits_mono}, the beam drastically damps the BAO features, and if opting to fit a wiggles/no-wiggles model which is entirely reliant on the presence of these features, then the perpendicular smoothing from the beam would render constraints on perpendicular parameters impossible. However, in the full-shape case as in \citet{Soares_Cunnington2021}, whilst the beam still introduces limitations, it will nevertheless leave some amplitude in the perpendicular power signal to fit to, thus providing potential for constraints.

Concerning the nuisance parameters, \autoref{fig:cross_auto_corner135_fit} shows how the auto-correlation exhibits strong degeneracies among some parameters, for example between the resolution $R$ and the overall amplitude $b_{\rm{\hi}}\overline{T}_{\rm{b}}$, which we consider as a single parameter. Such degeneracies are wholly or partly lifted in the cross-correlation, although others are introduced in the cross-correlation specific parameters, not shown in the figure. We found the reconstruction of $n_{\parallel}$ and $n_{\perp}$ (as introduced in \appref{sec:FG+noisesec} and \autoref{fg_comp}) does not provide single, well-defined values, nor does it seem that the best-fit values coincide for both of them, as posited when setting the fiduciary values.
\subsubsection{ $\sqrt{\Delta \chi^2}$ comparison and BAO detection}
\begin{figure}
    \includegraphics[width=\columnwidth]{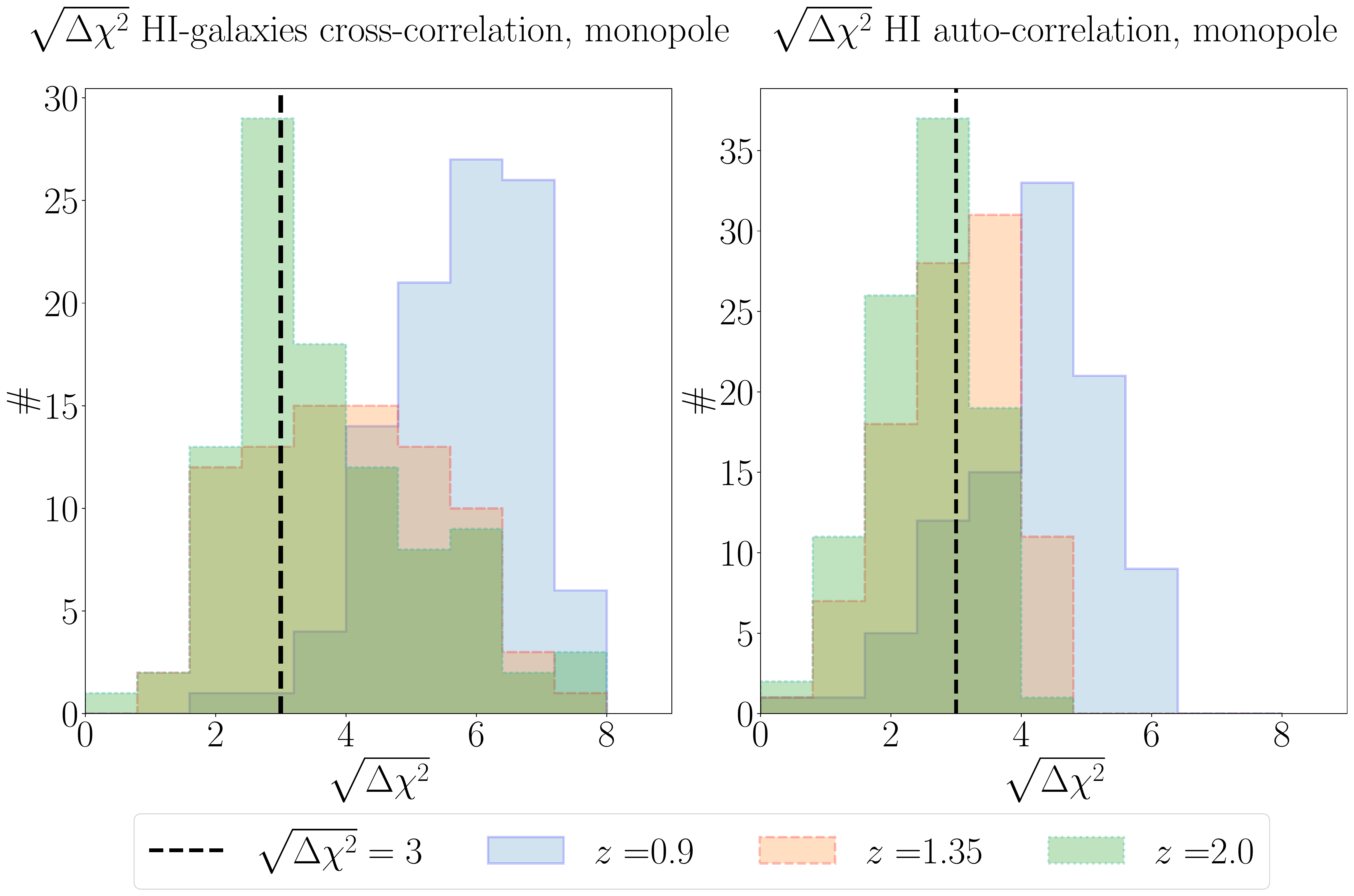}
    \caption{BAO detection significance in terms of $\sqrt{\Delta \chi^2}$ for both auto- and cross-correlation monopole at different redshifts. }
    \label{fig:deltachi_mono}
\end{figure}

To assess the significance of BAO detection, we follow the method presented in \citet{Villaescusa-Navarro:2016kbz}, which employs the $\chi^2$ for each best-fit posterior set.
We introduce the quantity
\begin{equation}
    \sqrt{\Delta \chi^2}=\sqrt{\chi_{\rm nw}^2-\chi_{\rm w}^2}\;, \label{eq:BAO_delta_detection}
\end{equation}
defining the BAO detection significance in terms of number of standard deviations, $\sigma$. We show the significance of BAO detection with the monopole for each redshift in both auto- and cross-correlation in \autoref{fig:deltachi_mono}. Cases for which a $P_{\rm nw}(k)$ template is a better fit than a $P_{\rm w}(k)$ cannot be shown in the histograms of \autoref{fig:deltachi_mono}, but we mark them as failed cases in  \autoref{tab:n_detected_bao_mono}.
\begin{table}
    \centering
    \caption{Comprehensive table of BAO detection in the monopole, distinguished by redshift and method. With failed we classify those simulations for which $\chi^2_{\rm nw}$ is smaller than $\chi^2_{\rm w}$.  }
    \begin{tabular}{l c c c c }
    \hline 
    Method & Redshift & \multicolumn{3}{c}{\% of realisations -- monopole} \\ 
    \cmidrule(l){3-5}
    & &
    $\geq 3\sigma\ (\leq \chi^2_{0.95})$ & $<3\sigma$ & failed  \\
    \hline
    auto & 0.90 & 78 (76)  & 19 & 3 \\ 
    cross & 0.90 & 99 (92)  & 1 & 0 \\
    auto & 1.35  & 50 (43) & 46 & 4 \\
    cross & 1.35 & 59 (31) & 38 &  3\\ 
    auto & 2.00 & 25 (14) & 71 & 4 \\
    cross & 2.00 & 50 (43) & 46 & 4 \\
    \hline 
    \end{tabular}
    \label{tab:n_detected_bao_mono}
\end{table}

We find that in the lowest redshift bin ($z=0.9$), both auto- and cross-correlation allow for a highly significant detection of the BAO, with the cross-correlation out-performing the auto-correlation. On the other hand, at higher redshift detections tend to move closer to the 3$\sigma$ threshold.

This method to assess the significance only evaluates the \textit{relative} difference between the wiggles and no-wiggles fitting template, and does not consider the intrinsic goodness of the fit.
Therefore, we also count how many of the $\chi^2$ in the subset of positively detected realisations are smaller or equal to the equivalent 95th percentile value of the variable (representing a $2\sigma$ confidence interval) and we quote the result in parentheses of \autoref{tab:n_detected_bao_mono}. 
We see from this that results for the cross-correlation achieve a better overall fit than the auto-correlation: occasional deviations from the trend could be a consequence of the loose assumptions on the compensation parameters priors, that could be avoided by stronger constrains. 
As expected, the number of null or below $3\sigma$ detections increases with redshift: this is due to a decreasing SNR, as discussed in the related \appref{sec:uncert_snr_section} and shown in \autoref{fig:SNR} (left panel). 
\subsection{Improving BAO detection with higher-order multipoles}
Previous works \citep[e.g.][]{Soares_Cunnington2021} showed how higher-order multipoles can improve parameter constraints in \hi\ intensity mapping experiments. In particular, recent results \citep{Kennedy:2021srz, Avila:2021wih} have demonstrated the benefit from including the quadrupole into \hi\ intensity mapping BAO detection attempts, albeit in the context of the configuration-space two-point correlation function. With this in mind, we investigate what benefits the quadrupole could provide to our Fourier-space BAO analysis. The formalism for the quadrupole calculation is outlined in \appref{Append_ref_pk}.

\begin{figure*}
    \centering
    \includegraphics[width=2.1\columnwidth]{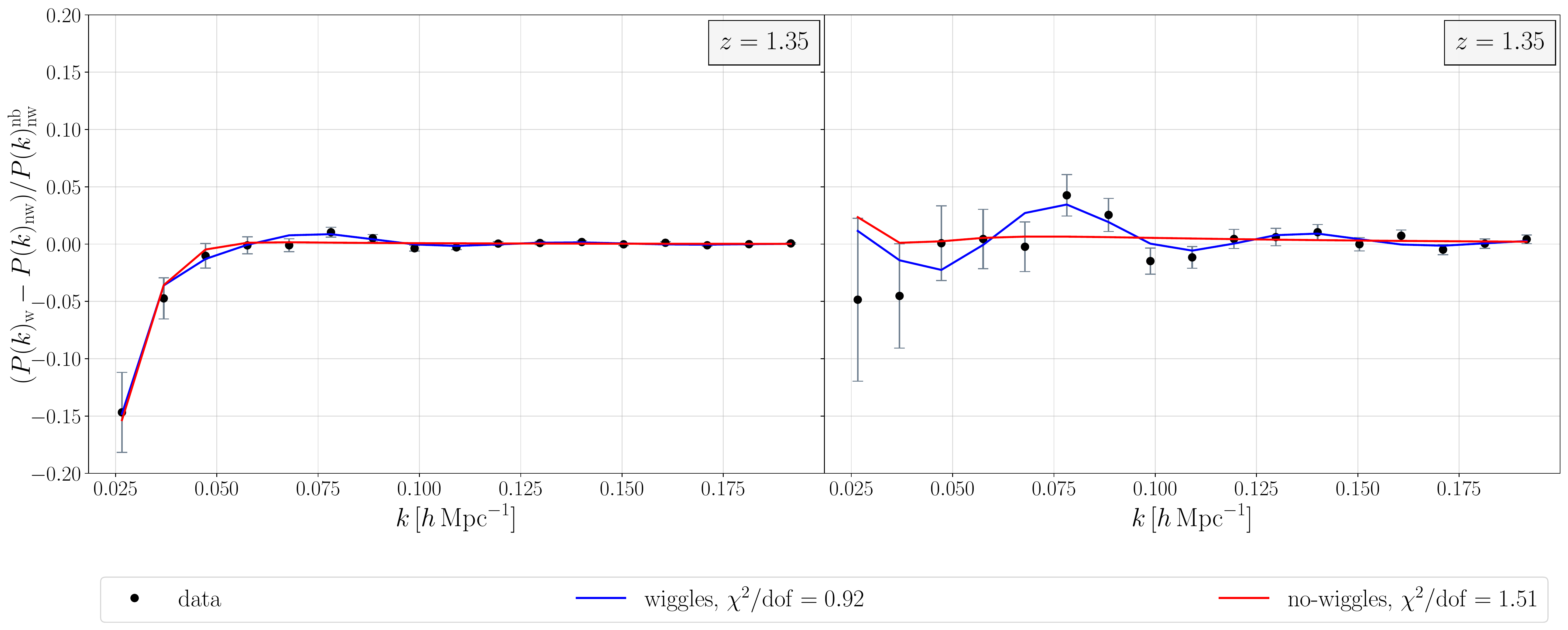}
    \caption{Example auto-correlation monopole (left) and quadrupole (right) power spectra for a single realisation at $z=1.35$.}
    \label{fig:stacked_fits_mono_quadru}
\end{figure*}

\autoref{fig:stacked_fits_mono_quadru} shows how the BAO features in the quadrupole (right panel) appear more robust with respect to the damping effects from the beam if compared with the monopole \hi\ auto-correlation (left panel). 
The different response of the higher-order multipoles to intensity mapping observational effects has been previously studied \citep{Blake_2019_auto_cross_errors_pipeline,Cunnington_Pourtsidou_Monopole} and this behaviour could mean that higher-order multipoles are of crucial importance to parameter constraints: more details about these features are included in the \appref{Appendix_mono_vs_quadru}. 
\begin{table}
    \centering
    \caption{Comprehensive table of BAO detection in the quadrupole, distinguished by redshift and method. With failed we classify those simulations for which $\chi^2_{\rm nw}$ is smaller than $\chi^2_{\rm w}$.  }
    \begin{tabular}{l c c c c }
    \hline 
    Method & Redshift & \multicolumn{3}{c}{\% of realisations -- quadrupole} \\
    \cmidrule(l){3-5}
    & &
    $\geq 3\sigma\ (\leq \chi^2_{0.95})$ & $<3\sigma$ & failed  \\
    \hline 
    auto & 0.90 & 77 (74)  & 21 & 2 \\
    cross & 0.90 & 89 (80)  & 10 & 1 \\
    auto & 1.35  & 54 (49) & 44 & 2 \\
    cross & 1.35 & 72 (57) & 27 &  1\\ 
    auto & 2.00 & 17 (14) & 74 & 9 \\
    cross & 2.00 & 17 (11) & 75 & 8 \\
    \hline 
    \end{tabular}
    \label{tab:n_detected_bao_quadrupole} 
\end{table}

Since the quadrupole appears more robust to damping effects from the beam by eye, we explored using the quadrupole alone in our BAO analysis. The number of positive detections is comparable to the monopole, except in the highest redshift bin where quadrupole detections are noticeably lower (see \autoref{tab:n_detected_bao_quadrupole} in comparison to \autoref{tab:n_detected_bao_mono}). Nevertheless, we found the quadrupole significance distributions always peaked at a lower number of $\sigma$ compared to the monopole. 
This is perhaps unsurprising given that the SNR is expected to decrease for higher-order multipoles. However, the relatively good signal from the quadrupole makes it a tantalising addition to a \hi\ intensity mapping BAO study and we therefore investigate a joint analysis including both monopole and quadrupole.

\subsubsection{Monopole-Quadrupole Joint Fit}

\begin{figure*}
    \centering
    \includegraphics[width=1.15\columnwidth]{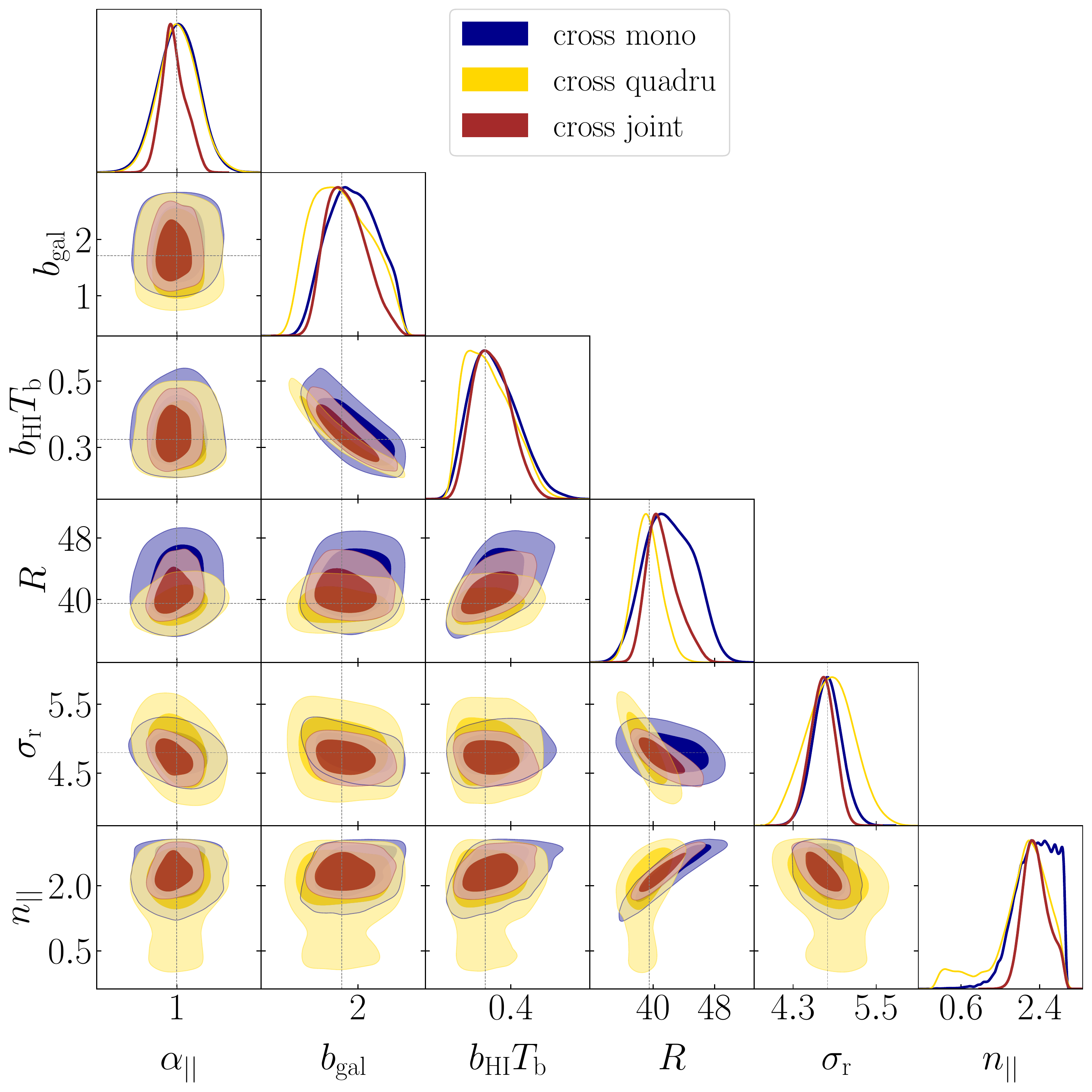}
    \includegraphics[width=0.8\columnwidth]{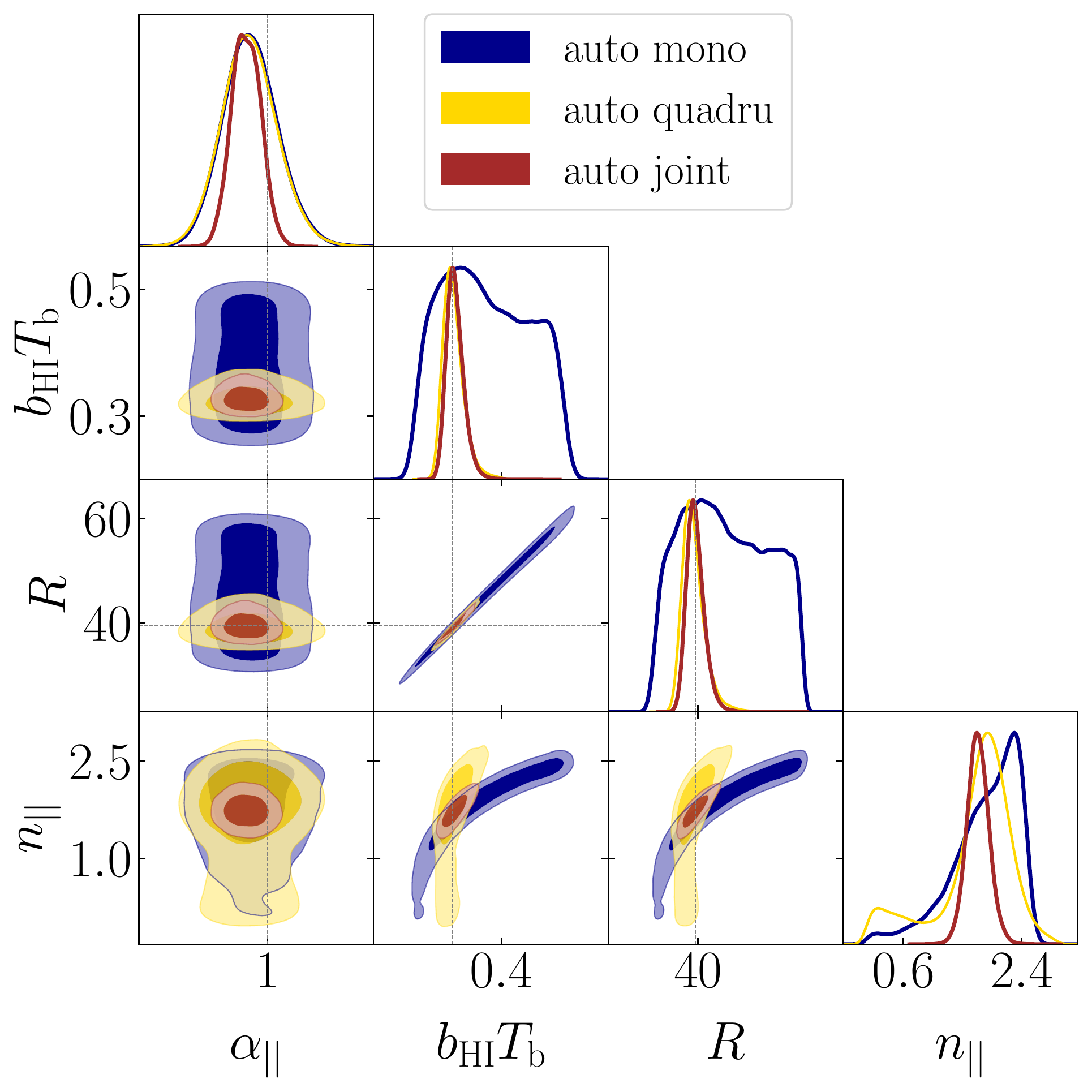}
    \caption{Example parameter posteriors ($z=$1.35) for cross-correlation (left-panel) and auto-correlation (right-panel) for a single chosen realisation. Shown are constraints using monopole-only (blue) quadrupole-only (yellow) and the joint fit case (red).
    }
   \label{fig:posteriors_sep_joint}
\end{figure*}

\begin{figure}
    \centering
     \includegraphics[width=\columnwidth]{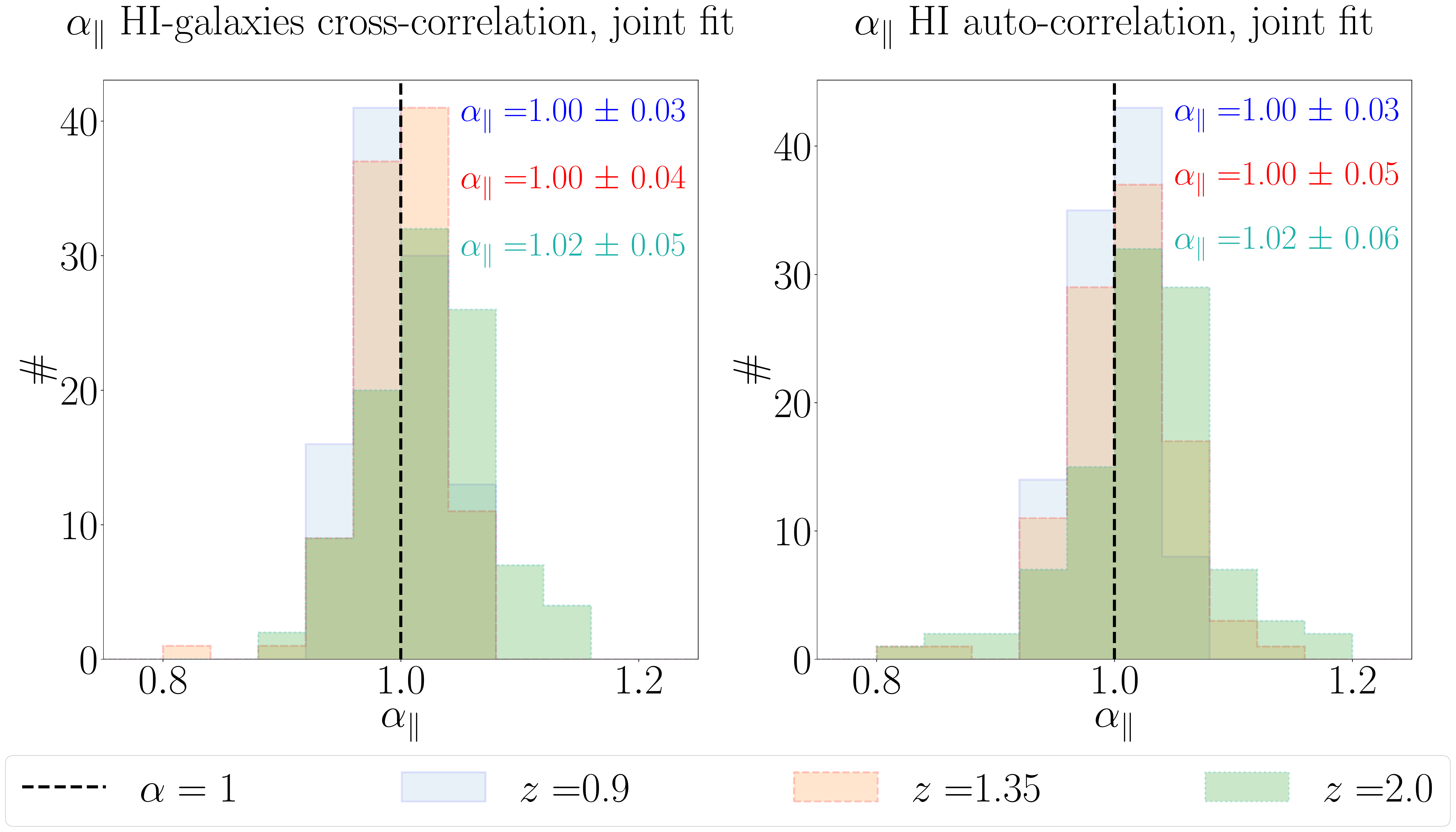}
    \caption{Distribution of the best fit $\alpha_\parallel$ 
    values for all realisations in the joint fit case, at different redshifts.} 
    \label{fig:alpha_joint}
\end{figure}

\begin{figure}
    \centering
    \includegraphics[width=\columnwidth]{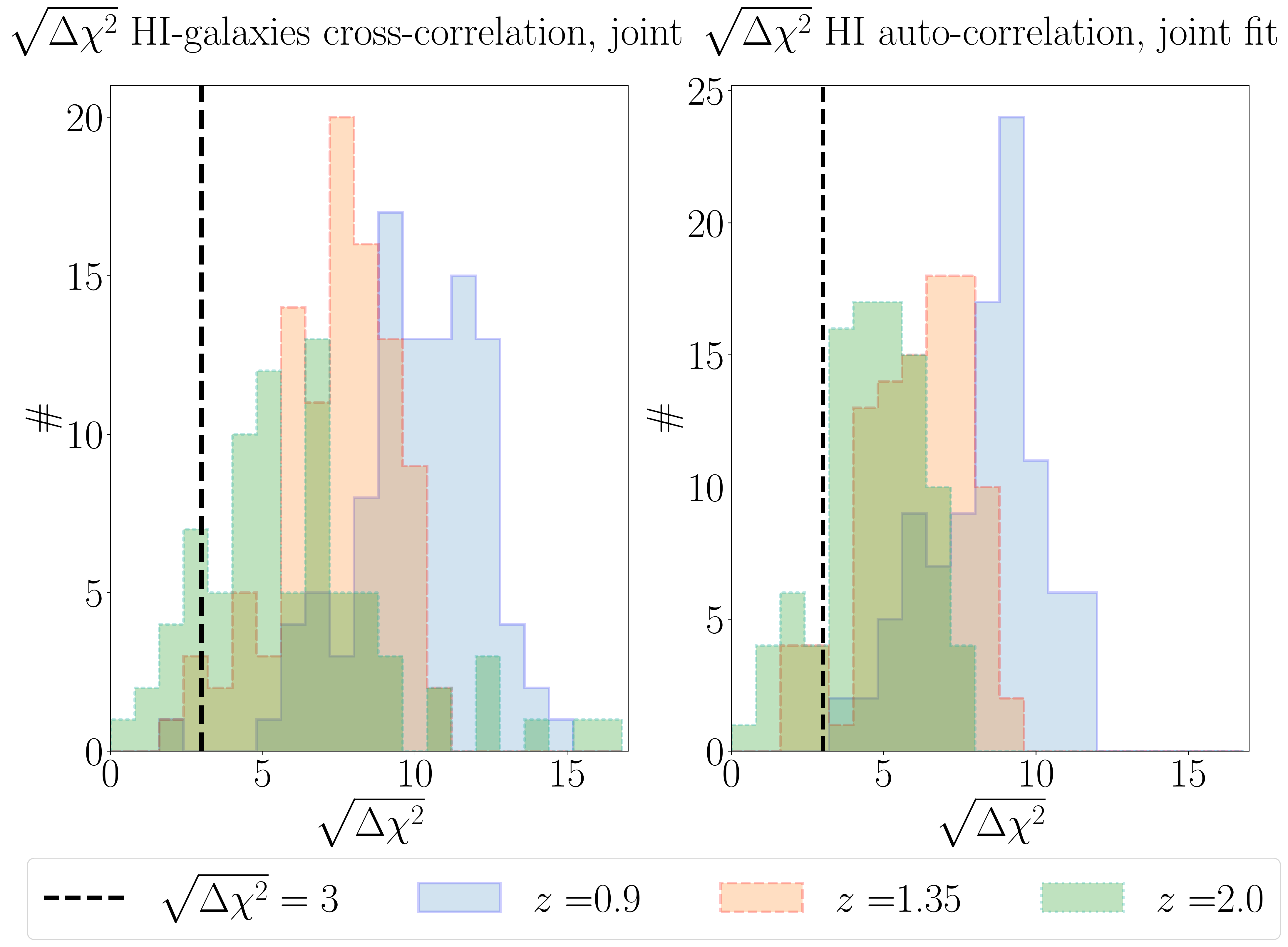}
    \caption{BAO detection significance in terms of $\sqrt{\Delta \chi^2}$ for monopole and quadrupole joint fit in auto- and cross-correlation at different redshifts.} 
    \label{fig:deltachi_joint}
\end{figure}

We now move to the case of the joint fit.
\autoref{fig:posteriors_sep_joint} shows a posterior distributions from one of the performed simulations for this joint fit approach for the auto-correlation (left panel, red contours) and the cross-correlation (right panel, red contours). To allow for an informed comparison with the individual monopole results, we include both the $P_{0}$ and $P_{2}$ separate fit posteriors (in blue and yellow respectively).
For the joint fit, we see a noticeable improvement in constraints for all parameters. Focusing on the cosmological AP parameters, \autoref{fig:alpha_joint} shows the distribution of best-fit $\alpha_\parallel$ obtained from our suite of simulations. We still see good $\alpha_\parallel$ constraints in both auto- and cross-correlation, again with a slight broadening/biasing of the distribution for higher redshifts. The $\alpha_\perp$ constraints still proved to be poor and remained prior dominated, even in a cross-correlation joint fit: having observed such a result in three different methods, we would conclude that single-dish intensity mapping will struggle to be a competitive probe of $\alpha_\perp$ for instruments with SKA-like beam sizes.
\autoref{fig:deltachi_joint} demonstrates the boost in successful detections from the joint fit approach (in comparison with the $P_{0}$-only results of \autoref{fig:deltachi_mono}) and that the detection significance achieved in cross-correlation is even higher compared to auto-correlation analyses. We also summarise the number of detections for the joint fit in \autoref{tab:n_detected_bao_joint}, which highlights an interesting contrast between the high number of positive detections---meaning that the increase of information plugged into the likelihood contributes to a better model discrimination---and the number of detections below the 95th percentile threshold of the $\chi^2$, quoted in parentheses---indicating a quite poor intrinsic agreement between data and model.

There could be a number of reasons for this. We recall that the analysis exploits two different goodness-of-fit indicators: the posteriors behaviour, more nuanced; and the $\chi^2$, summarising the result in a single quantity, necessary to establish the BAO detection significance as defined in \autoref{eq:BAO_delta_detection}. Part of the explanation could rely on intrinsic limitations of using $\chi^2/\text{dof}$ as a measure of goodness-of-fit, in particular concerning the correct evaluation of the number of degrees of freedom (in our case, the sum of the $P_{0}$ and $P_{2}$ data vector lenghts minus the number of fitting parameters in either models), becoming increasingly non-trivial for complex models with priors \citep{Andrae:2010gh}. From this point of view, by performing a joint fit, we have complicated this calculation: a proper analysis for the dof is beyond the scope of this paper, and we instead focus more on the results from the detections significance and the parameter constraints from the MCMC analysis.

Another possible cause, more tightly related to the present model and data, could be due to the role of the covariance term being introduced in the joint fit case (\autoref{eq:likelihood_general}, see also \autoref{fig:heatmap_cross_auto_09}).
Firstly, the joint fit has intrinsically more degrees of freedom: the higher this value, the narrower the $\chi^2$ probability density function, thus less forgiving with high $\chi^2$ values. In addition, if the covariance term is non-negligible, the resulting $\chi^2$ is larger than the sum of $\chi^2_{{P_0}}$ and $\chi^2_{{P_2}}$, resulting in an additional penalty.
It follows that the joint fit allows a better wiggles vs no-wiggles discrimination because of a higher SNR resulting from the terms in the likelihood all positively contributing to the result (see \autoref{eq:snr_def} and the bottom-most panel of \autoref{fig:SNR}). However, for the same reason, it gives rise to a lower goodness-of-fit, its effect being analogous to an underestimation of the uncertainties. 

\begin{table}
    \centering
    \caption{Comprehensive table of BAO detection in the monopole-quadrupole joint fit, distinguished by redshift and method. With failed we classify those simulations for which $\chi^2_{\rm nw}$ is smaller than $\chi^2_{\rm w}$.} 
    \begin{tabular}{l c c c c }
    \hline 
    Method & Redshift & \multicolumn{3}{c}{\% of realisations -- joint} \\
    \cmidrule(l){3-5}
    & &
    $\geq 3\sigma\ (\leq \chi^2_{0.95})$ & $<3\sigma$ & failed  \\
    \hline 
    auto & 0.90 & 98 (62)  & 0 & 2 \\ 
    cross & 0.90 & 99 (49)  & 1 & 0 \\
    auto & 1.35  & 91 (40) & 8 & 1 \\
    cross & 1.35 & 95 (38) & 4 & 1\\ 
    auto & 2.00 & 79 (26) & 15 & 6 \\
    cross & 2.00 & 69 (14) & 11 & 20 \\
    \hline 
    \end{tabular}
    \label{tab:n_detected_bao_joint}
\end{table}
\section{Discussion and conclusions} \label{sec:disc}
We collect here our conclusions on the study of the \hi\ auto- and \hi\,galaxy cross-correlation power spectra Legendre multipoles as a tool for the detection of the baryon acoustic oscillations (BAO) in future, state-of-the-art radio and optical/near-infrared cosmological experiments.

Our data sets are based on log-normal simulations in the redshift interval $z\in[0.9,\,2.0]$, which are generated from an input power spectrum based on a vanilla $\Lambda$CDM cosmology. To make the input signal as realistic as possible the input power spectrum includes a boost from linear RSD and non-linear effects, produced using \texttt{halofit}. Furthermore, we include observational effects from the radio telescope beam, thermal noise, and astrophysical foregrounds---the latter then cleaned using a blind PCA approach.
We summarise our main conclusions below:
\begin{enumerate}
    \item BAO detection appears most likely feasible for both auto- and cross-correlation at the lowest redshift we tested ($z=0.9$), where $P_{0}$ and $P_{2}$ have a large majority of realisations above the 3$\sigma$ threshold, e.g.\ $78\%$ for the \hi\ auto-correlation monopole, increasing to $99\%$ in \hi\,galaxy cross-correlation. At higher redshift, we find that the cross-correlation remains to have a generally higher significance than the auto-correlation. However, the number of realisations exceeding the $3\sigma$ threshold decreases with redshift, likely caused by the increasing beam size, higher noise contributions, and a reduction in galaxy density affecting the cross-correlation benefit. Such clear-cut redshift-dependent `hierarchy' is consistent with previous works \citep{Villaescusa-Navarro:2016kbz}, where tighter constraints were found from the intensity mapping radial power spectrum at low redshift, except in the lowest redshift bin at $z\sim0.5$. (Note that this is outside our adopted redshift range.) We made the choice of a maintaining a survey size with consistent physical dimensions at every redshift, $l_\text{x},\,l_\text{y},\,l_\text{z} = \{2400,\,2400,\,600\}\,\mathrm{Mpc}\,h^{-1}$. However, in reality a fixed angular size e.g.\ $\sim4000\,\text{deg}^2$ would be surveyed, giving larger physical dimensions at higher redshifts, perhaps providing the possibility of more optimistic higher redshift constraints. Approximately speaking, we expect the uncertainty on a power spectrum measurement to decrease as $1/\sqrt{V_\text{sur}}$ where $V_\text{sur}$ is the volume of the survey. The estimated sky fractions for each of our redshifts were $f_\text{sky}=0.12,\,0.07,\,0.05$. Since $V_\text{sur}$ is directly proportional to $f_\text{sky}$, this suggests that uncertainties on our highest redshift data could decrease by a factor of $1/\sqrt{2}$ if assuming the same $f_\text{sky}$ as the lowest redshift data. This is because the $f_\text{sky}$ at the lowest redshift is approximately twice that of the highest redshift. This overly basic calculation offers some encouragement for the higher redshift case, but we leave a more detailed investigation into this for future work.
    
    \item Under certain conditions, the quadrupole alone could efficiently detect the BAO (see \autoref{tab:n_detected_bao_quadrupole}) and seems to display more prominent BAO-like features (\autoref{fig:stacked_fits_mono_quadru}), demonstrating more robustness to the damping caused by the dominant beam effects. Similar results were found in \citet{Kennedy:2021srz} and \citet{Avila:2021wih} for the intensity mapping correlation function, the reasons for which were investigated in \citet{Cunnington_Pourtsidou_Monopole}, a discussion we also outline in \appref{Appendix_mono_vs_quadru}. Essentially this is caused by the down-weighting of beam-dominated modes in the quadrupole. Despite the apparent increased robustness to the beam, we found the quadrupole detection significance distribution peaks at lower values than the corresponding monopole. The reason for this is the decreased SNR in the quadrupole and is demonstrated in \autoref{fig:SNR}.
    \item We find an excellent increase in detection significance when we perform a joint fit for the monopole and quadrupole. \autoref{fig:deltachi_joint} shows the majority of realisations achieved a $\sigma\sim 10$ detection, much higher when compared to the monopole-only results in \autoref{fig:deltachi_mono}. However, paradoxically we found only a minority of their $\chi^2$ is below the 95th percentile threshold. The good levels of SNR for the joint fit (see \autoref{fig:SNR}) could be interpreted as an underestimation of the uncertainties, which could justify the result, together with an important role of the covariance that, as an additional term, biases the $\chi^2$ towards higher values.
    At the same time, we observe an important reduction of the marginalised contour sizes of the posteriors for the join-fit (\autoref{fig:posteriors_sep_joint}), suggesting an improved capability to constrain parameters.
   
    \item The ability to resolve the `wiggled' power spectrum with respect to the broad-band counterpart depends on the contribution of two main factors: the telescope smoothing and the error bar size, which, in turn, depends on both the telescope smoothing and the noise sources that appear in \autoref{cross_err} and \ref{auto_err_tot}. On the signal side, the stronger the damping, the flatter the power spectrum: it follows that distinguishing between the BAO and no-wiggles power spectra becomes increasingly difficult.
    Regarding measurement uncertainties, the damping, growing with the redshift, may partially contribute to reducing the size of the error bars, but not as efficiently as the suppression of the power spectrum amplitude.
    Not only is the denominator of the signal-to-noise ratio conditioned by the damping factor: we also have to consider it is given by a sum of different terms. As a result, the SNR decreases with $z$ and it is not necessarily constant in the $k$-interval corresponding to the BAO scale. Since larger error bars translate into smaller $\chi^2$ values, lower SNR also contribute to make broad-band and `wiggled' power spectra less distinguishable. 
    We can conclude that, however complicated the concurrence of factors,  mitigating the transverse beam is a primary necessity, as shown by the better performance of the cross-correlation: the contribution of the radial smoothing, whose typical size is almost constant for every redshift, is limited and acts only in the highest-$k$ modes of our region of interest.
    
    \item Parameter prediction provides mixed outputs. Focusing on the cosmological $\alpha$ parameters, we find $\alpha_\parallel$, connected to the measurement of $H(z)$ along the line of sight, can be well constrained with the cross-correlation offering a slight improvement in the best-fit value distributions. We found some small biasing begins to appear at our highest redshift, as shown by the $\alpha_\parallel$ constraints displayed in \autoref{fig:alpha_mono} and \autoref{fig:alpha_joint}, although this is still comfortably within the $1\sigma$ bounds in all cases. On the other hand, probably due to the heavy transverse smoothing effect, we have an almost complete loss of information on $\alpha_{\perp}$, whose posteriors are for the most part were found to be uniformly distributed throughout the prior interval. Significantly, the AP factors are robust against correlations with the other quantities in the explored parameter space. Cross-correlation breaks or relieves some of the stronger degeneracies between astrophysical and instrumental parameters we observe in the auto-correlation, but tends to introduce other analogous correlation among its own nuisance parameters.

    \item Foreground removal can be successfully realised via the PCA algorithm, provided that the undesired excessive subtraction of cosmological power at scales below $k<0.1\,h\,\mathrm{Mpc}^{-1}$ be corrected by a compensation window. For our scopes, the adopted compensation model relies on purely phenomenological considerations (see \citet{Cunnington_Pourtsidou_Monopole} and \citet{Soares_Cunnington2021} for a detailed discussion of the \hi\ auto-correlation fit with and without compensation models).
    
    \item We observe an excellent agreement between the variance of our data set and the assumption of Gaussian analytic uncertainties; we confirm the existence of correlation among multipoles when the beam and foreground cleaning is included (see \appref{Appendix_uncert}) as shown in \citet{Soares_Cunnington2021}.
    
    \item Given SKA and \textit{Euclid} future fields of view, surveys with wider transverse size than in our mocks could no doubt be considered. In this work our simulations were mainly aimed at a nearer future experiment such as that done by SKA's precursor MeerKAT.
    Larger volumes would inevitably reduce error bar size and lead to an improved detection of BAO with both methods and multipole expressions. We have been particularly careful to keep our mock data depth along the grid's $\mathrm{z}$-direction minimal, to avoid the issue of evolving parameters with redshift. However, a sufficient implementation of a redshift weighting could be considered \citep[see e.g.][]{Mueller:2017pop,Ruggeri_Percival_extended_BAO_2018}, thus allowing for further increase survey size leading to an additional improvement in constraints.
\end{enumerate}

Generally our results should be interpreted as optimistic for using \hi\ intensity mapping BAO for constraining cosmology. The results can be seen as an extension of \citet{Villaescusa-Navarro:2016kbz}, finding similar conclusions in that the large beam from an SKA or MeerKAT survey erodes most transverse information. Where \citet{Villaescusa-Navarro:2016kbz} exclusively uses the radial 1D power spectrum to avoid this issue, we stick to the 3D spherically averaged power spectrum, which we find can still recover good radial information, and also significantly benefits from the inclusion of the quadrupole, which naturally down-weights beam dominated modes. One thing to explore in future work would be using clustering $\mu$-wedges \citep[as investigated in][]{Avila:2021wih}, as a more tailored approach to avoiding regions particularly affected by systematics.

One potential benefit from the multipole approach is that it provides an alternative method for treating non-linear effects. For the three-dimensional $P(\bm{k})$, non-linear effects can be largely avoided by ignoring large $k$ values. Doing this in the radial $P(k_\parallel)$ does not guarantee an avoidance of non-linear effects since even small $k_\parallel$ can still be affected by the non-linearities of physical
small scales. Furthermore, isolating or modelling observational effects could potentially be more troublesome for the radial $P(k_\parallel)$ method, as shown in \citet{Matshawule:2020fjz} and \citet{spinelli2021skao_blind_challenge}.
In any case, it is important for alternative approaches to be available for pursuing precision cosmology. 
Hence our results, which show a successful implementation of the three-dimensional $P(\bm k)$ for measuring BAO, boosted by cross-correlations and quadrupole inclusion, are encouraging since this could prove to be a regime where non-linear and observational effects can be more optimally treated.

\section*{Acknowledgements}
The authors wish to thank Alkistis Pourtsidou and David Bacon for useful discussions at the beginning of this project. We also thank Santiago Avila, Phil Bull, Isabella Carucci and Paula Soares for useful feedback. AR and SCa acknowledge support from the `Departments of Excellence 2018-2022' Grant (L.\ 232/2016) awarded by the Italian Ministry of University and Research (\textsc{miur}). SCu is supported by a UK Research and Innovation Future Leaders Fellowship, grant MR/S016066/1 (PI: Alkistis Pourtsidou) and also acknowledges support from the STFC grant ST/S000437/1. SCa also acknowledges support by \textsc{mur} Rita Levi Montalcini project `\textsc{prometheus} -- Probing and Relating Observables with Multi-wavelength Experiments To Help Enlightening the Universe's Structure', for the early stages of this project, and from the `Ministero degli Affari Esteri della Cooperazione Internazionale (\textsc{maeci}) -- Direzione Generale per la Promozione del Sistema Paese Progetto di Grande Rilevanza ZA18GR02.

We acknowledge the use of open source software \citep{scipy:2001, Hunter:2007, mckinney-proc-scipy-2010, numpy:2011,  GetDist}. Some of the results in this paper have been derived using the healpy and HEALPix package.

\section*{Data Availability}
The data underlying this article will be shared on reasonable request to the corresponding author.




\bibliographystyle{mnras}
\bibliography{main_bib}


\appendix
\section{Power Spectrum Formalism} \label{Append_ref_pk}
\subsection{Signal}

We estimate the \hi\ auto- and cross-power spectra for our simulated data by first taking the Fourier transform of the \hi\ intensity map temperature fluctuations $\delta T_\textrm{\hi}(\mathbf{x})\,{=}\,T_\textrm{\hi}(\mathbf{x})\,{-}\,\overline{T}_\text{b}$ and the galaxy number fluctuations $\delta_\text{g}(\mathbf{x})\,{=}\,(n_\text{g}(\mathbf{x})\,{-}\,\overline{n}_\text{g})/\overline{n}_\text{g}$. 
\begin{equation}
    \tilde{F}_\textrm{\hi}(\mathbf{k})=\sum_{\mathbf{x}} \delta T_\textrm{\hi}(\mathbf{x}) \exp (i \mathbf{k}{\cdot}\mathbf{x})\,,
\end{equation}
\begin{equation}
    \tilde{F}_\text{g}(\mathbf{k})=\sum_{\mathbf{x}} \delta_\text{g}(\mathbf{x}) \exp (i \mathbf{k}{\cdot}\mathbf{x})\,.
\end{equation}
%
The power spectra can then be estimated with
%
\begin{equation}
    \hat{P}_\textrm{\hi}(\mathbf{k}) = V_\text{cell}\lvert\tilde{F}_\textrm{\hi}(\mathbf{k})\rvert^2\,,
\end{equation}
\begin{equation}
    \hat{P}_\textrm{\hi,\text{g}}(\mathbf{k}) = V_\text{cell}\operatorname{Re}\left\{\tilde{F}_\textrm{\hi}(\mathbf{k})\cdot \tilde{F}^{*}_\text{g}(\mathbf{k})\right\}\ .
\end{equation}
%
where $V_\text{cell}\,{=}\,l_\text{x}l_\text{y}l_\text{z}/N_\text{grid}^3$. These power spectra are then spherically averaged and, in the case of the quadrupole, simultaneously weighted by the Legendre multipole (see \appref{sec:LegMultipole}). The power spectra are modelled as
%
\begin{align}
P_\textrm{\hi\,\hi}(k,\mu,z)&=b_{\text {\rm\hi}}^2(z) \overline{T}_{\text b}^2(z) P_{\text {\rm m}}(k,z)\left[1+\frac{f(z)}{b_{\text{\hi}}(z)}\mu^2\right]^2\;,\label{eq:Pauto}\\
P_\textrm{\hi\,g}(k,\mu,z)&=b_{\text {\rm\hi}}(z)\overline{T}_{\text b}(z)b_{\rm g}(z)P_{\rm m}(k,z)\nonumber\\
\phantom{=}&\times\left[1+\frac{f(z)}{b_{\text{\hi}}(z)}\mu^2\right]\left[1+\frac{f(z)}{b_{\text{g}}(z)}\mu^2\right]\;,\label{eq:Pcross}
\end{align}
where $b_{\text {\rm{\hi}}}$ and $b_{\text g}$ are respectively the \hi\ and galaxy linear biases, $\overline{T}_\mathrm{b}(z)$ is the mean \hi\ brightness temperature, $f(z)\approx\Omega_{\text m}(z)^\gamma$ is the growth rate, with $\gamma\simeq0.545$ the growth index, and $P_{\text m}$ is the (linear) matter power spectrum. In \autoref{eq:Pcross}, we have assumed that any cross-correlation coefficient between the \hi\ and galaxies is unity. In reality this is unlikely to be the case and a coefficient may exhibit some more complex scale dependence \citep{Wolz:2015ckn,Anderson:2017ert}. Strictly speaking, there should also be some normalisation by the amplitude of cosmological fluctuations $\sigma_8$ in the above formalism. Whilst for some probes the degeneracy between $\sigma_8$ and the mean brightness temperature $\overline{T}_\text{b}$ could cause issues \citep{Castorina:2019zho}, BAO probes are less affected by this degeneracy, since they are not overly sensitive to the normalisation. Therefore we do not to consider its inclusion.

From the expressions above, the units of the auto-correlation power spectrum 
$[h^{-3}\, \mathrm{Mpc^3\, mK^2}]$ and of the cross-correlation power spectrum $[h^{-3}\, \mathrm{Mpc^3\, mK}]$ can be easily inferred; furthermore, the reader will understand why, while $b_{\text g}(z)$ alone can be used as a fitting parameter, the quantity we look for in \hi\ measurements is the combined pair $b_{\text {\rm\hi}}(z) \overline{T}_{\text b}(z)$, introducing $b_{\text \hi}$ as a fixed number in the RSD factor. The adopted models for the \hi\ bias and the $\overline{T}_\mathrm{b}(z)$ factor are
\begin{align}
    b_{\rm{\hi}}&=0.904+0.135(1+z)^{1.696},\\
    \overline{T}_{\rm{b}}&=190 \frac{H_{0}(1+z)^2}{H(z)} \Omega_{\rm{\hi}}(z) \,h\,\rm{mK},
\end{align}
with $\Omega_{\rm{\hi}}(z)=4 \times 10^{-4}(1+z)^{0.6}$. These models are from \citet[][eqs. 2 and A3]{Villaescusa-Navarro:2016kbz}, whilst the $b_{\text g}(z)$ factor is recovered from the interpolation of the data shown in \citet[][Table 3]{EuclidVII:2020}.

\subsection{Observational Effects}\label{sec:FG+noisesec}
\subsubsection{Shot Noise}\label{sec:shot_noise}
There are expected sources of contamination to the cosmological signal, both for \hi\ intensity mapping and galaxy surveys. For the latter, shot-noise is the standard form of measurement noise, which we emulate by Poisson-sampling a finite number of galaxies. For the former, shot-noise is expected to be low due to the broad integration of signal \citep{Spinelli:2019smg}, which in our simulation we avoid by using a biased form of the raw output density field from the log-normal generator.

\subsubsection{21-cm Foregrounds}
Since foregrounds dominate the \hi\ signal, they require cleaning, a process we emulate by adding realistic radio foreground simulations into our data, then applying the PCA cleaning algorithm to the contaminated data. The method assumes that the first $N_\text{fg}$ principal components from the intensity mapping data frequency covariance matrix will contain the highly correlated and dominant foregrounds \citep{Liu:2011ih,Masui:2012zc}. By removing this small number of dominant components we should largely remove the foreground contamination but simultaneously exhibit some signal loss to the \hi\ modes most degenerate with the foregrounds---typically large radial modes \citep{Switzer:2015ria}. This should sufficiently emulate any signal loss to the \hi\ which is an important limitation for \hi\ intensity mapping experiments. Many other foreground removal algorithms are available, the most successful of which are blind techniques, e.g.\ FastICA \citep{Wolz:2013wna}, GMCA \citep{Carucci:2020enz}, GPR \citep{Soares:2021ohm}, KPCA \citep{Irfan:2021bci}, to name a few. 

To simulate radio foregrounds, we resort to the Global Sky Model provided by \texttt{PyGSM} \citep{PyGSM1,PyGSM2}, a software relying on the \texttt{healpy} and \texttt{HEALPix} packages\footnote{\href{http://healpix.sourceforge.net}{healpix.sourceforge.net}} \citep{Healpy1,Healpy2}, which produces full-sky maps covering the emission from $0.01$ to $100\,\mathrm{GHz}$ based on many real data sets, corresponding to a number of physical sources, in particular the cosmic microwave background, the synchrotron, warm and cool dust emissions, and free-free processes. We convert the full sky maps from spherical coordinates to a data box suitable for our log-normal simulations and add these into our simulation. The sky region identified for the foreground simulation bares little impact on the success of the foreground removal (assuming no polarisation leakage or other complex systematics) as identified in \citet{Cunnington:2020njn}. We chose a region centred on the galactic plane and assuming the high amplitude spectra remain smooth from this region, which is guaranteed from our simulation, foreground removal was still possible.

We find that to clean the signal $N_\text{fg}=3$ is sufficient for all redshift bins and for both cross- and auto-correlation, consistent with previous simulation-based studies \citep{Alonso:2014dhk, Cunnington:2020njn}. More aggressive cleans are expected to be required when in the presence of systematic issues from e.g.\ polarisation leakage or beam effects which we do not investigate in this work \citep[see e.g.][]{Carucci:2020enz,spinelli2021skao_blind_challenge}. It is expected that these issues should be avoidable either by modelling the systematics using information from the scanning strategy \citep{McCallum:2021jsv} or exquisite instrument calibration which should hopefully be achievable with a full SKAO survey. 
As already mentioned, and as expected from former literature \citep[][]{Cunnington_2019_foregrounds,Cunnington_Pourtsidou_Monopole}, the foreground removal procedure affects the cosmological signal on the largest scales, therefore we introduce a phenomenological compensation term $B_{\text{fg}}(k,\mu)$. This is based upon a function aimed at modelling signal suppression due to the finite volume surveyed and can be given by \citep{Bernal_2019}
\begin{multline}
B_{\text{vol}}(k,\mu)=\left(1-\exp\left \lbrace -\left(\frac{k}{k_{\parallel,\min}}\right)^2 \mu^2\right \rbrace\right)\\ \times\left(1-\exp\left \lbrace- \left(\frac{k}{k_{\perp,\min}}\right)^2 \left(1- \mu^2 \right)\right \rbrace\right)\,.
\label{vol_comp}
\end{multline}
Incorporating a modified version of this into the power spectrum model has been shown to mitigate the effects from foreground cleaning \citep{Soares_Cunnington2021,Cunnington:2020wdu} and can also be applicable in higher-order statistics for the case of the \hi\ intensity mapping bispectrum \citep{Cunnington:2021czb}. Following \citet{Soares_Cunnington2021}, we define the foreground compensation term as
\begin{multline}
B_{\text{fg}}(k,\mu)=\left(1-\exp\left \lbrace -\left(\frac{k}{n_{\parallel} k_{\parallel,\min}}\right)^2 \mu^2\right \rbrace\right)^{q/2}\\ \times\left(1-\exp\left \lbrace- \left(\frac{k}{n_{\perp} k_{\perp,\min}}\right)^2 \left(1- \mu^2 \right)\right \rbrace\right)^{q/2}. 
\label{fg_comp}
\end{multline}
In both \autoref{vol_comp} and \ref{fg_comp} we define $k_{\parallel,\min}=2 \pi/l_{\rm z}$ and $k_{\perp,\min}=2 \pi/\sqrt{l_{\rm x}^2+l_{\rm y}^2}$, based on the box size. We will henceforth denote the product of the terms as $B_{\textrm{vol,fg}}(k,\mu)$ for the sake of brevity. Nonetheless, the foreground window additionally depends on the pair of free parameters $n_{\parallel}$, $n_{\perp}$ and applies to the sole $\delta_\text{\hi}$ field, so that the exponent $q$ is either equal to 1 in the cross-correlation or 2 in the auto-correlation case.

In this work we have no strong theoretical motivations to justify the values of $n_{\parallel}$ and $n_{\perp}$, whose optimal value shall be found by the MCMC, their fiduciary values having been determined `by eye'. The only remarks we have are that both parameters are in the order of unity and that the radial direction seems to influence the correction much more than the transverse direction. This latter consideration is particularly true for the monopole, the quadrupole appears slightly more sensitive to the perpendicular correction.
We found the average best-fit values from visual inspection for $n_{\parallel}$ and $n_{\perp}$ at each of our three redshifts $z=0.9,1.35,2.0$ to be: for the cross-correlation, $\{1.9,1.9\}$, $\{1.85,1.85\}$, and $\{1.9,1.9\}$; for the auto-correlation, $\{1.45,1.45\}$, $\{1.45,1.45\}$, and $\{1.6,1.6\}$. Interestingly, we find that in cross-correlation they tend to be larger than for auto.

\subsubsection{Smoothing from Beam and Redshift Uncertainty}

We model the SKAO observation using the most elementary telescope model, a Gaussian beam with transverse (redshift-dependent) smoothing scale, given by 
\begin{equation}
    R=\frac{r(z)\,\theta_{\text{\rm FWHM}}}{2\sqrt{2 \ln2}}\,,
\end{equation}
where $r(z)$ is the radial comoving distance to redshift $z$ and $\theta_{\text{\rm FWHM}}$ is the beam angular resolution, which given SKA features, can be approximated as $0.8\, (1+z) \, \text{deg}$. More sophisticated analyses, including beam sidelobes, can be found in \citet{Matshawule:2020fjz} and \citet{spinelli2021skao_blind_challenge}.

Similarly, to simulate future observations in the optical/near-infrared with \textit{Euclid} or \textit{Roman}, we introduce a redshift uncertainty factor $\sigma_r$, again modelled by Gaussian smoothing but this time in the radial direction. Following e.g.\ \citet[][Eqs 74 and 75]{EuclidVII:2020}, we write $\sigma_r=c\,(1+z)/H(z)\,\sigma_{0,z}$ with $\sigma_{0,z}=0.001$.

Hence, the beam induces a smoothing in the transverse direction for the \hi\ field, i.e.
\begin{equation}
    \delta_{\text{\rm{\hi,obs}}}({\bm k}_\perp,k_\parallel,z)={\rm e}^{-k_\perp^2R^2/2}\delta_\text{\rm{\hi}}({\bm k}_\perp,k_\parallel,z)\;,
    \label{eq:delta_auto}
\end{equation}
and the redshift uncertainty induces a smoothing in the radial direction for the galaxies field, namely
\begin{equation}
    \delta_{\mathrm{g,obs}}({\bm k}_\perp,k_\parallel,z)={\rm e}^{-k_\parallel^2\sigma_r^2/2}\delta_\mathrm{g}({\bm k}_\perp,k_\parallel,z)\;,
    \label{eq:delta_cross}
\end{equation}
where the presence or absence of the subscript `obs' denotes either the observed field or the underlying cosmological field. Therefore, the observed power spectra, comprehensive of volume and foreground-subtraction compensation windows (defined in \autoref{fg_comp}), and with the RSD included in the $\mu$ dependence, respectively read
\begin{equation}
P_\textrm{\hi\,\hi,obs}(k,\mu,z)={\rm e}^{-k^2R^2(1-\mu^2)}P_\textrm{\hi\,\hi}(k,\mu,z) B_{\text{fg,vol}}(k,\mu)
\label{obs_full_auto}
\end{equation}
and
\begin{multline}
P_\textrm{\hi\,g,obs}(k,\mu,z)={\rm e}^{-k^2/2 [R^2(1-\mu^2) +\sigma_r^2 \mu^2]}\\\times P_\textrm{\hi\,g}(k,\mu,z) B_{\text{fg,vol}}(k,\mu) \;.
\label{obs_full_cross}
\end{multline}
As shown in the paper, the role of the smoothing term is fundamental for the BAO
with the magnitude of the term depending on the value of the beam size $R$ and on the considered scale.
The benefits coming from the cross-correlation are the reduction of the transverse damping to the square root of the analogous auto-correlation term.

\subsubsection{Thermal Noise}\label{sec:thermal_noise}

A potentially large source of contamination for \hi\ intensity mapping comes from random thermal noise fluctuations caused by the instrument itself. This is well approximated with a flat (scale-invariant) power spectrum defined by \citep{Santos:2015gra}
\begin{equation}
    P_{\text{th}}(z) =
    \frac{4\pi\,c\,f_{\text{sky}} [(1 + z)\,r]^2}
    {2\,t_{\text{tot}}\, \nu_{\text{21}}\,N_{\text{dish}}\,H(z)}\,T_{\rm sky}^2(z)\;.\label{thermo_1}
\end{equation}
In this work we use $N_{\text{dish}}=200$, ${t_{\text{tot}}}=10\,000\,\mathrm{hours}$, and
\begin{equation}
    T_{\rm sky}(z) = T_{\text{inst}} +60\left(\frac{\nu}{300\,\mathrm{MHz}}\right)^{-2.5}\;,\label{thermo_2}
\end{equation}
with $T_{\text{inst}}=25\,\mathrm{K}$. This noise therefore has some redshift dependence through the redshifted frequency, $\nu(z)$. These chosen values are approximately consistent with a planned SKAO intensity mapping experiment \citep{SKA_Redbook_2020}. 

With such values and for the scales of our interest, the thermal noise is expected to contaminate only the highest $k$ bins of cosmological signal at intermediate and high redshift.
Given the thermal noise power spectrum, we can easily produce a Gaussian random field describing this instrumental noise to be added to the \hi\ maps. 

\subsection{Legendre Multipole Expansion}\label{sec:LegMultipole}

We can now decompose the power spectrum in multipoles, i.e.\ projecting it on the basis of the Legendre polynomials $\mathcal{L}_{\ell}$, thus having
\begin{equation}
P(k,\mu,z)= \sum_{\ell} P_{\ell}(k,z) \mathcal{L}_{\ell}(\mu)\;. 
\label{eq:Legendre_decomp}
\end{equation}
Then, the $\ell$th multipole $P_{\rm X,\ell}(k,z)$ of the power spectrum, with $X=\{\textrm{\hi\,\hi},\,\textrm{\hi\,g}\}$, can be obtained by integrating over $\mu$ such that for the auto-correlation we have
\begin{equation}
P_{\textrm{\hi\,\hi},\ell}(k,z)=\frac{2\ell+1}{2}\int_{-1}^{1}\de \mu\ P_\textrm{\hi\,\hi,obs}(k,\mu,z) \mathcal{L}_{\ell}(\mu)\;, \label{eq:ell_autoHI}\end{equation} 
and for the cross-correlation we have
\begin{equation}
P_{\textrm{\hi\,g},\ell}(k,z)=\frac{2\ell+1}{2}\int_{-1}^{1}\de \mu\ P_\mathrm{\hi,\text g,obs}(k,\mu,z) \mathcal{L}_{\ell}(\mu)\;. \label{eq:ell_cross}\end{equation}
For $\ell=0$ and $\mathcal{L}_{0}=1$  we have the monopole, for $\ell=2$ and $\mathcal{L}_{2}=\left(3 \mu^2-1\right)/2$ we have the quadrupole; we also remind the reader that, when the AP-test is applied, neither the $\mathcal{L}_{\ell}$ terms nor the $\de\mu$ term are rescaled with $\alpha_{\parallel}$ and $\alpha_{\perp}$, whereas all other $\mu$ terms are redefined accordingly.
\begin{figure*}
    \includegraphics[height=0.8\textwidth]{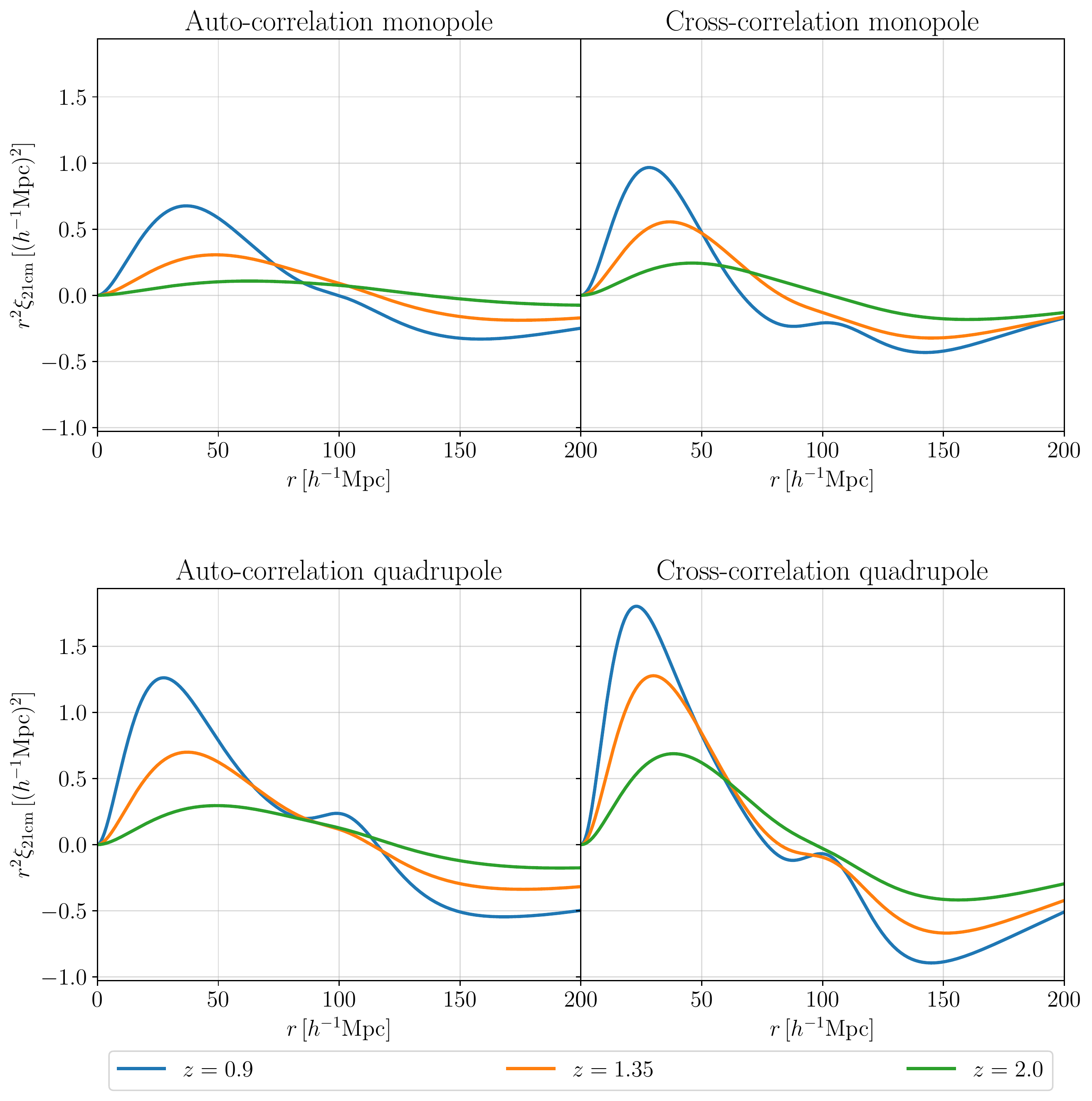}
    \caption{Correlation functions for the monopole and the quadrupole in both auto-correlation and cross-correlation. The displayed functions take into account all the instrumental, physical and phenomenological contributions to the power spectrum}
    \label{fig:correlationfunction}
\end{figure*}
 
It is common for BAO experiments to be analysed in configuration space \citep[e.g.][]{Bautista:2020ahg}, where the two-point correlation function can be found with the following relation:
\begin{equation}
    \xi_{X,{\rm obs}}(r,z)=\int\de k\,\frac{k^2}{2\pi^2}\frac{\sin(kr)}{kr}P_{X,\ell}(k,z)\;.
    \label{eq:2pcf}
\end{equation}
We plot the different behaviours of the function for the auto- and cross-correlation cases in \autoref{fig:correlationfunction}, showing the smearing of the excess probability peak at $r\approx105\,\mathrm{Mpc}\,h^{-1}$ as a function of the instrumental effects, of the redshift, methods, and Legendre multipole.\footnote{The algorithm used to calculate the two-point correlation function is available from \href{https://github.com/JoeMcEwen/LOG_HT}{github.com/JoeMcEwen/LOG\_HT}, which is in turn based on \href{https://jila.colorado.edu/~ajsh/FFTLog/}{\texttt{FFTLog}}, originally developed for FORTRAN.} For work dedicated to BAO detection from the \hi\ intensity mapping correlation function we refer the reader to \citep{Kennedy:2021srz,Avila:2021wih}.

In this work, we have simulated our mocks onto a Cartesian grid and can therefore make a perfect plane-parallel (flat-sky) assumption. However, for future \hi\ intensity mapping surveys with wider sky coverage, increasing consideration would need to be given to wide-angle effects \citep{Yoo:2010ni,Bonvin:2011bg,Challinor:2011bk,2018JCAP...03..019T,2018JCAP...10..032T,Castorina:2017inr,Blake:2018tou}. These would not only impact the modelling of RSD \citep{Castorina:2019hyr} but also the anisotropic observational effects caused by foreground cleaning and the telescope beam. We thus highlight that future survey analysis may require to implement some extended modelling, beyond what we present. These large-scale effects can also be circumvented with other techniques such as probing BAO with harmonic-space (angular) power spectra (see e.g.\ \citet{DES:2021esc}.
\section{Gaussian uncertainties and Signal-to-Noise Ratio}
Errors can be assigned to the power spectra based on either the data themselves, e.g.\ via jackknifing or bootstrapping methods, or calculated \textit{a priori} with analytic formulas.
We use this second approach: the ansatz of normally distributed density contrasts underlying our simulations has as a main consequence the absence of correlation among modes out of the diagonal, as shown in \autoref{fig:heatmap_cross_auto_09}.
In addition, this assumption will allow for an explicit calculation of the magnitude of the error bars, which will be the subject of the following subsections.

\subsection{Formalism}\label{Appendix_uncert}
Our calculations follow \citep[][]{Soares_Cunnington2021}, in turn mostly deriving from \citep[][]{Bernal_2019}. Although using somewhat different notations and conventions, we also mention as a general reference  \citet{Blake_2019_auto_cross_errors_pipeline}, which makes a thorough examination of general auto- and cross-correlation power spectra definitions, correction terms and associated errors.

We define
%
\begin{align} \label{cross_auto_err_1}
    \mathcal{A}(k,\mu,z)=&\left[P_\textrm{g\,g}(k,\mu,z)+P_{\text{shot}}(k,\mu,z)\right]\\
    &\times\left[P_\textrm{\hi\,\hi}(k,\mu,z)+P_{\text{th}}(k,\mu,z)\right]\;,\\
    \mathcal{B}(k,\mu,z)=&\left[P_\textrm{\hi\,g}(k,\mu,z)\right]^2\;,\\
    \mathcal{C}(k,\mu,z)=&\left[P_\textrm{\hi\,\hi}(k,\mu,z)+P_{\text{th}}(k,\mu,z)\right]^2\;,
\end{align}
where the complete power spectra definitions can be found in \appref{Append_ref_pk}. Therefore, we can compose our full monopole-quadrupole covariance matrix, under the hypothesis that they are all diagonal in $k-k^\prime$, as 
\begin{align}
    C_{\ell,\ell^\prime}^{\text{auto}}=&\frac{(2 \ell+1)(2 \ell^\prime+1)}{N_{\bm k}} \int_{-1}^{1}\de\mu\;\mathcal{C}(k,\mu,z)\,\mathcal{L}_{\ell}(\mu)\,\mathcal{L}_{\ell^\prime}(\mu)\;.   \label{auto_err_tot}\\
    C_{\ell,\ell^\prime}^{\text{cross}}=&\frac{(2 \ell+1)(2 \ell^\prime+1)}{2 N_{\bm k}}\nonumber\\
    &\times\int_{-1}^{1}\de\mu\;\left[\mathcal{A}(k,\mu,z)+\mathcal{B}(k,\mu,z)\right]\,\mathcal{L}_{\ell}(\mu)\,\mathcal{L}_{\ell^\prime}(\mu)\;.  \label{cross_err}
\end{align}

Note that, for the sake of readability and brevity, instrumental smoothing, various prefactors, and additional corrections are left implicit, summarised under the $k$, $\mu$, and $z$ dependence of the power spectra terms. The term
$P_{\text{shot}}=1/n_{\text{gal}}(z)$ is the shot noise, depending on galaxy counts and modelled interpolating the data shown in \citep[Table 3]{EuclidVII:2020} with the formula exposed in the Eq.~(113) of the same paper, whereas $P_{\text{th}}$ is the antenna thermal noise, described by \autoref{thermo_1} and \ref{thermo_2}. Both noise power spectra are functions of $\mu$ and $k$, next to the default dependency on $z$, because both fields undergo smoothing.
From these definitions, uncertainties on data in the separate fit case are just the square root of the terms for which $\ell=\ell^\prime$  and the explicit calculation can be performed by assigning to the free parameters their fiducial values.

Lastly, concerning the denominator $N_{\bm k}$, viz.\ the number of independent modes available in the observed volume, this can be expanded as follows: $N_{\bm k}=k^2\, \Delta k\, V_{\rm sim} /{2 \pi^2}$, where $\Delta k$ is set equal to $k_{\text{min}}=2 \pi/L$, with $L$ smallest side of the simulation box \citep[as in][]{Camera_Fonseca_2020_OIII}, and $V_{\rm sim}$ is the total volume of the simulation.

\subsection{Validation on Data} \label{sec:uncert_snr_section}

\subsubsection{Full Covariance Matrix}
\begin{figure}
    \centering
    \includegraphics[width=0.9\columnwidth]{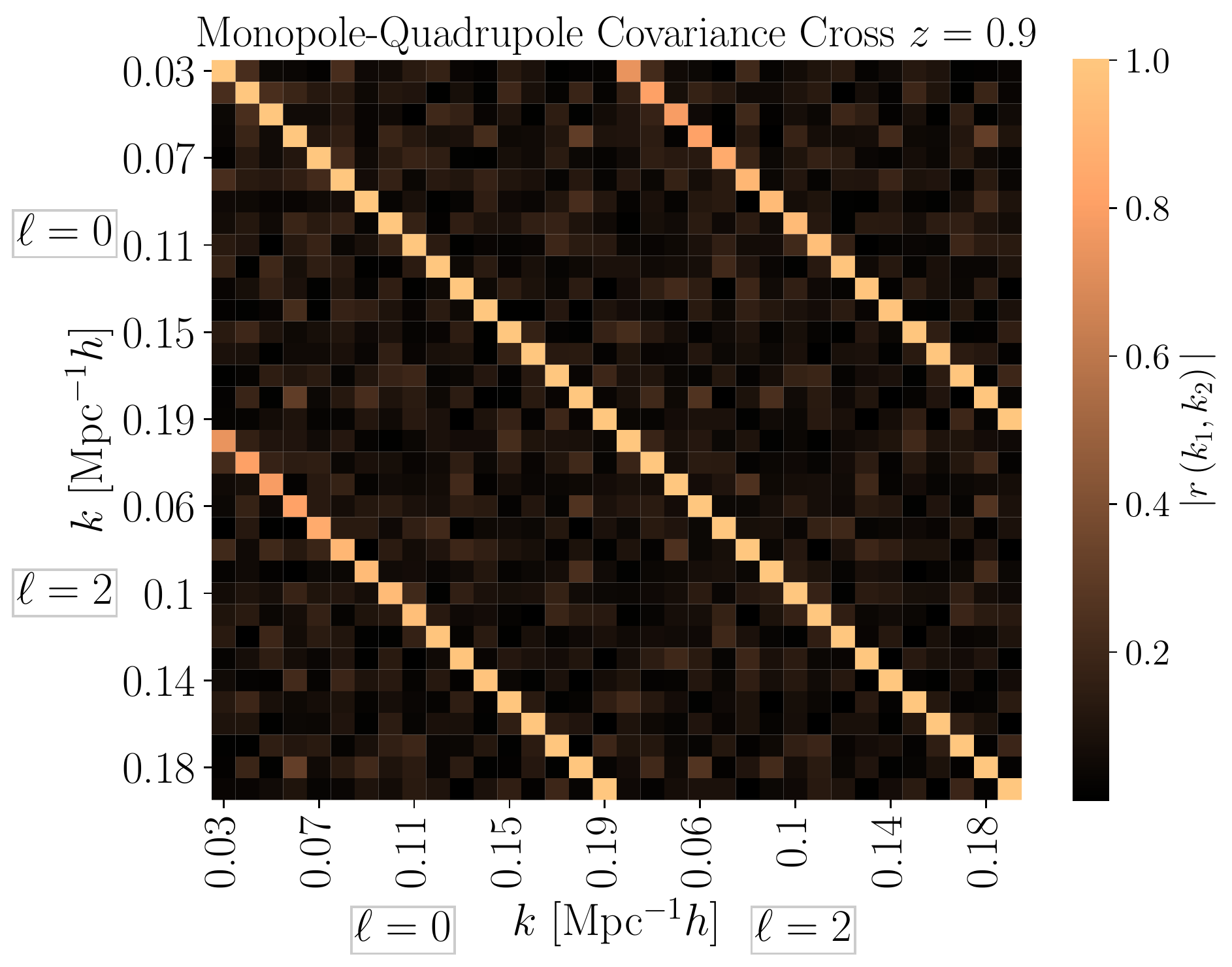}
    \includegraphics[width=0.9\columnwidth]{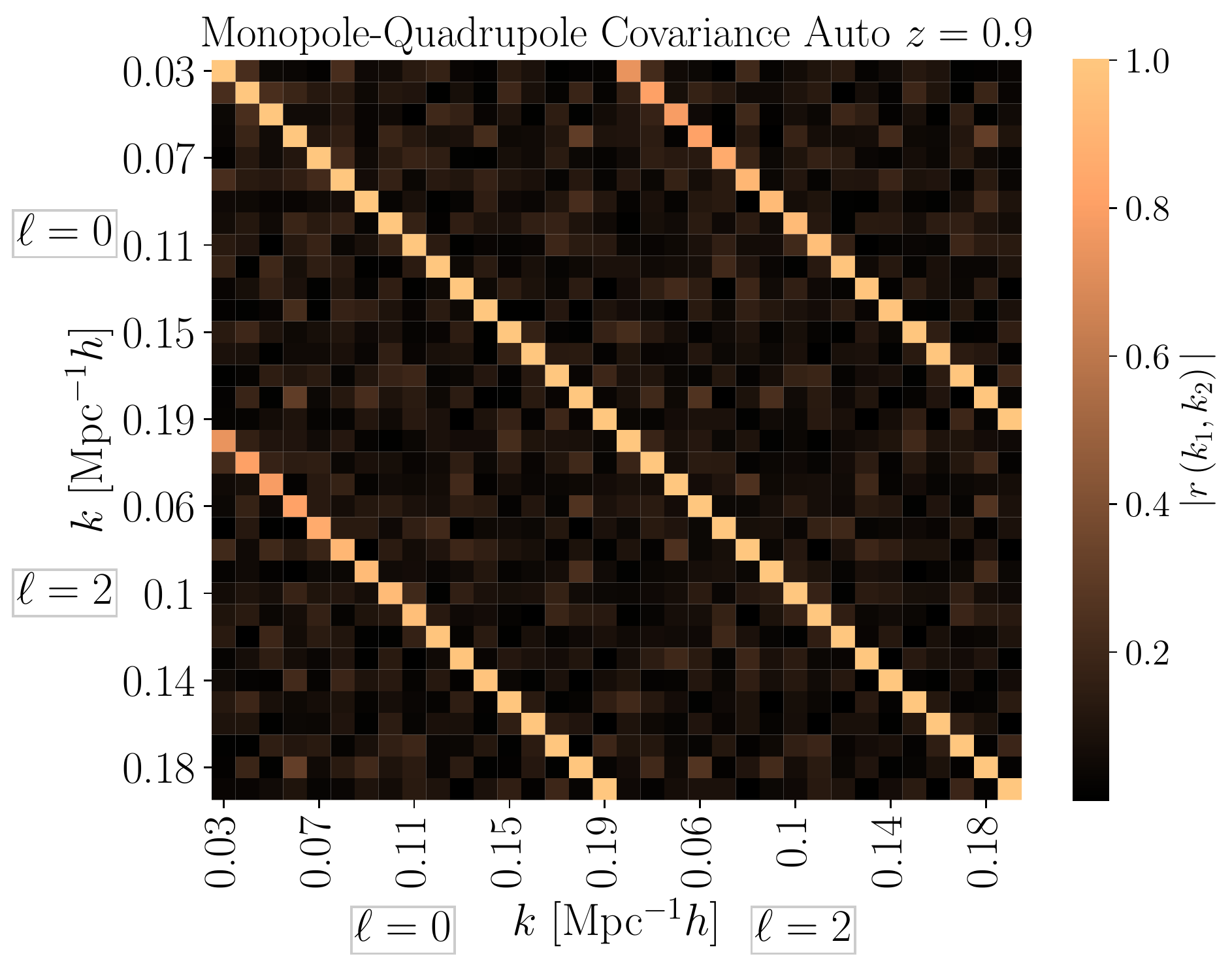}
    \caption{Covariance matrix at $z=0.9$ for the cross- (left) and auto-correlation (right) monopole and quadrupole power spectra. }\label{fig:heatmap_cross_auto_09}
\end{figure}

The absolute value of the linear correlation coefficient, itself reading
\begin{equation}
r_{\ell,\ell^\prime}(k_{i},k_{j}) =\frac{C_{\ell,\ell}(k_{i},k_{j})}{\sqrt{C_{\ell,\ell^\prime}(k_{i},k_{i})C_{\ell,\ell^\prime}(k_{j},k_{j})}}\;, \label{eq_linear_corr}
\end{equation}
is defined in terms of the covariance $C$ between any pair of modes, $k_{i}$ and $k_{j}$, and of multipoles, $\ell$ and $\ell^\prime$. The heatmap in \autoref{fig:heatmap_cross_auto_09} is calculated at the lowest redshift value, $z=0.9$, where non-linearities induced by the \texttt{halofit} prescription may most significantly affect BAO-scale modes. 
The covariance matrix is subdivided in four blocks, each one being diagonal: the blocks along the principal diagonal are the monopole-monopole and quadrupole-quadrupole covariances, whereas those on the secondary diagonal represent the symmetric monopole-quadrupole cross-covariance term.
The latter appears non-negligible: its role was already pointed out and explicitly calculated in a simplified case in \citet{Soares_Cunnington2021}, where it is shown that the presence of a beam---or of any $\mu$-dependent term---breaks the orthogonality of the multipoles.

\subsubsection{Gaussian Assumption}
\begin{figure}
    \centering
    \includegraphics[width=0.9\columnwidth]{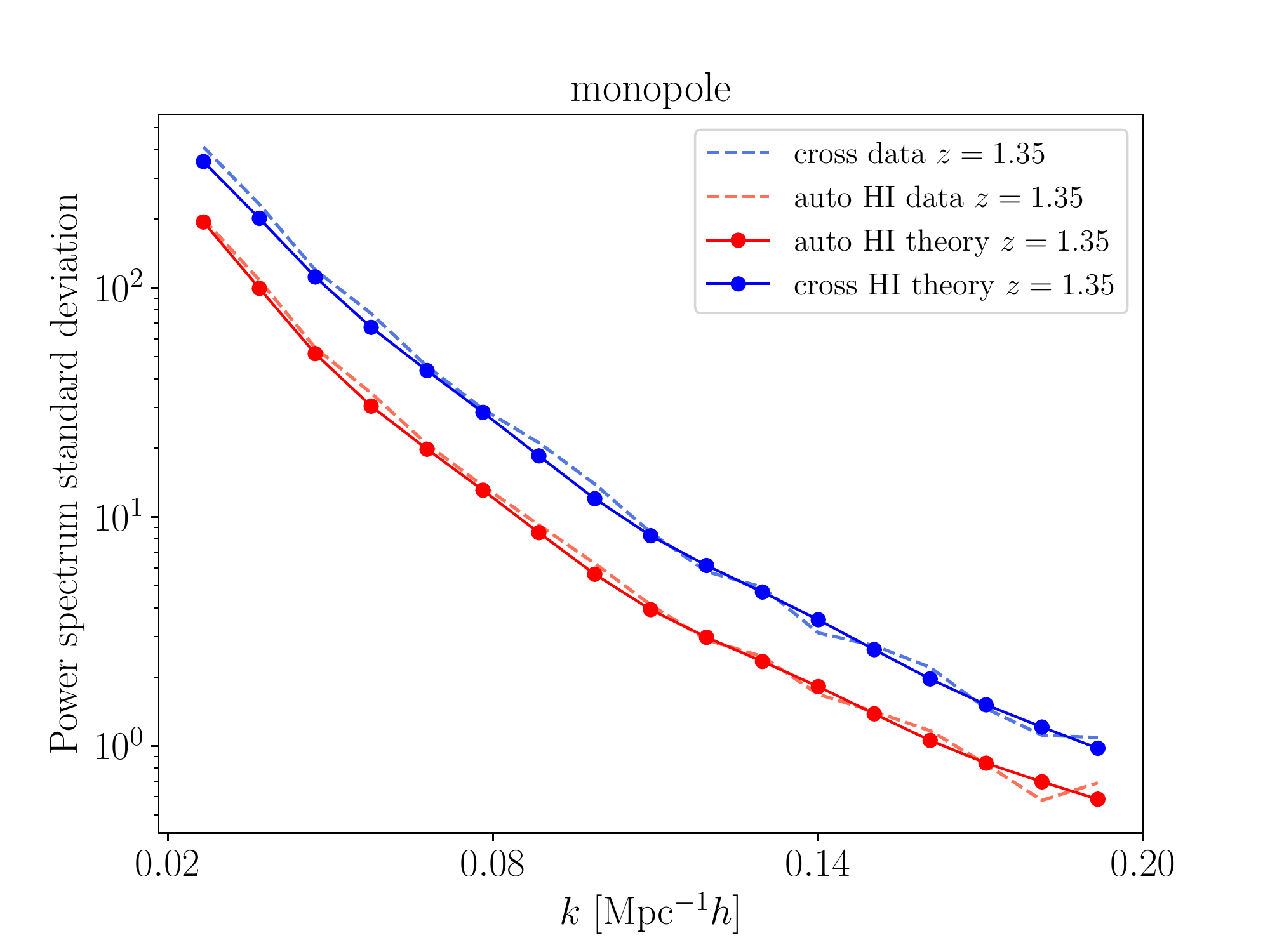}
    \includegraphics[width=0.9\columnwidth]{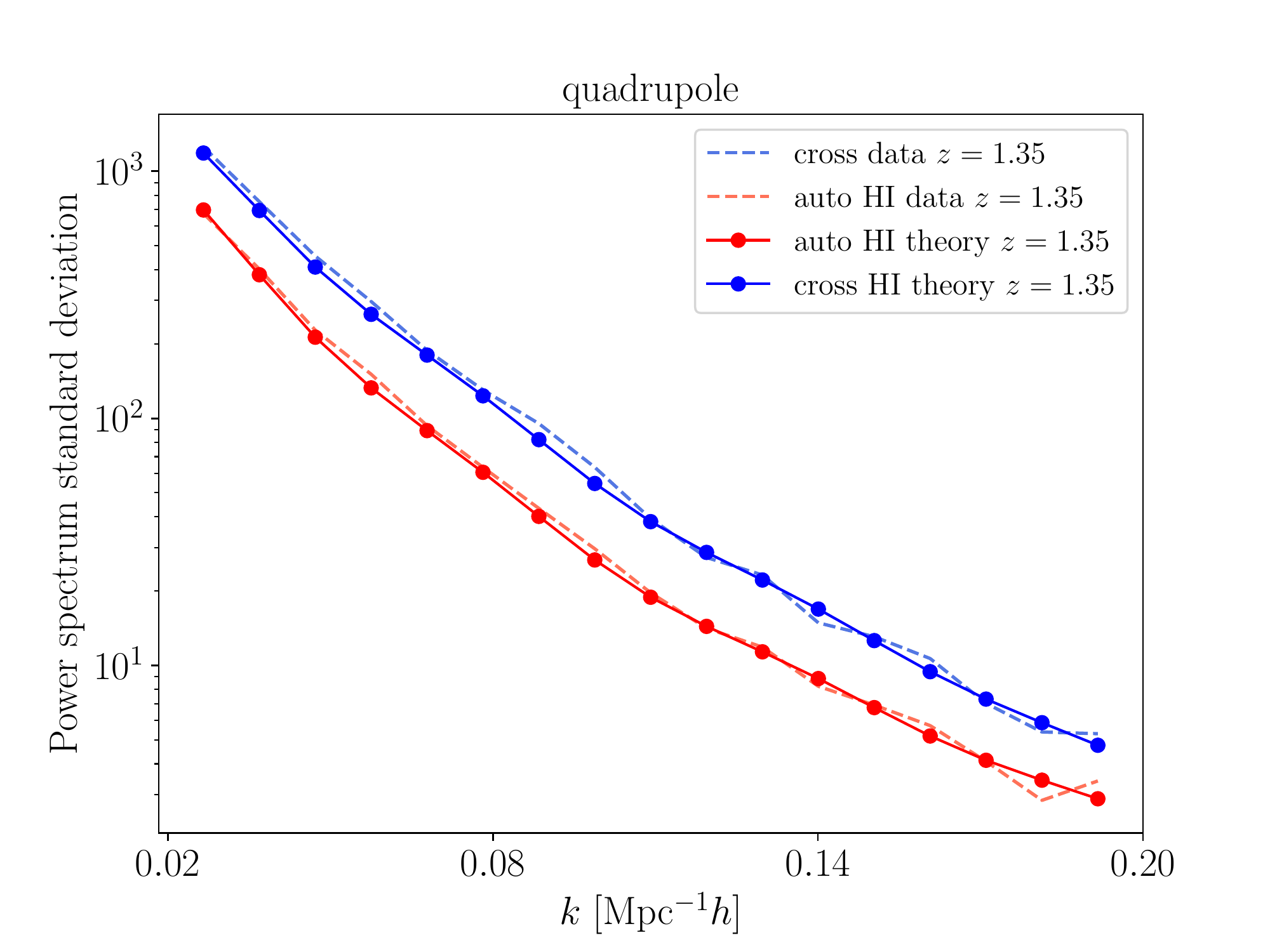}
    \includegraphics[width=0.9\columnwidth]{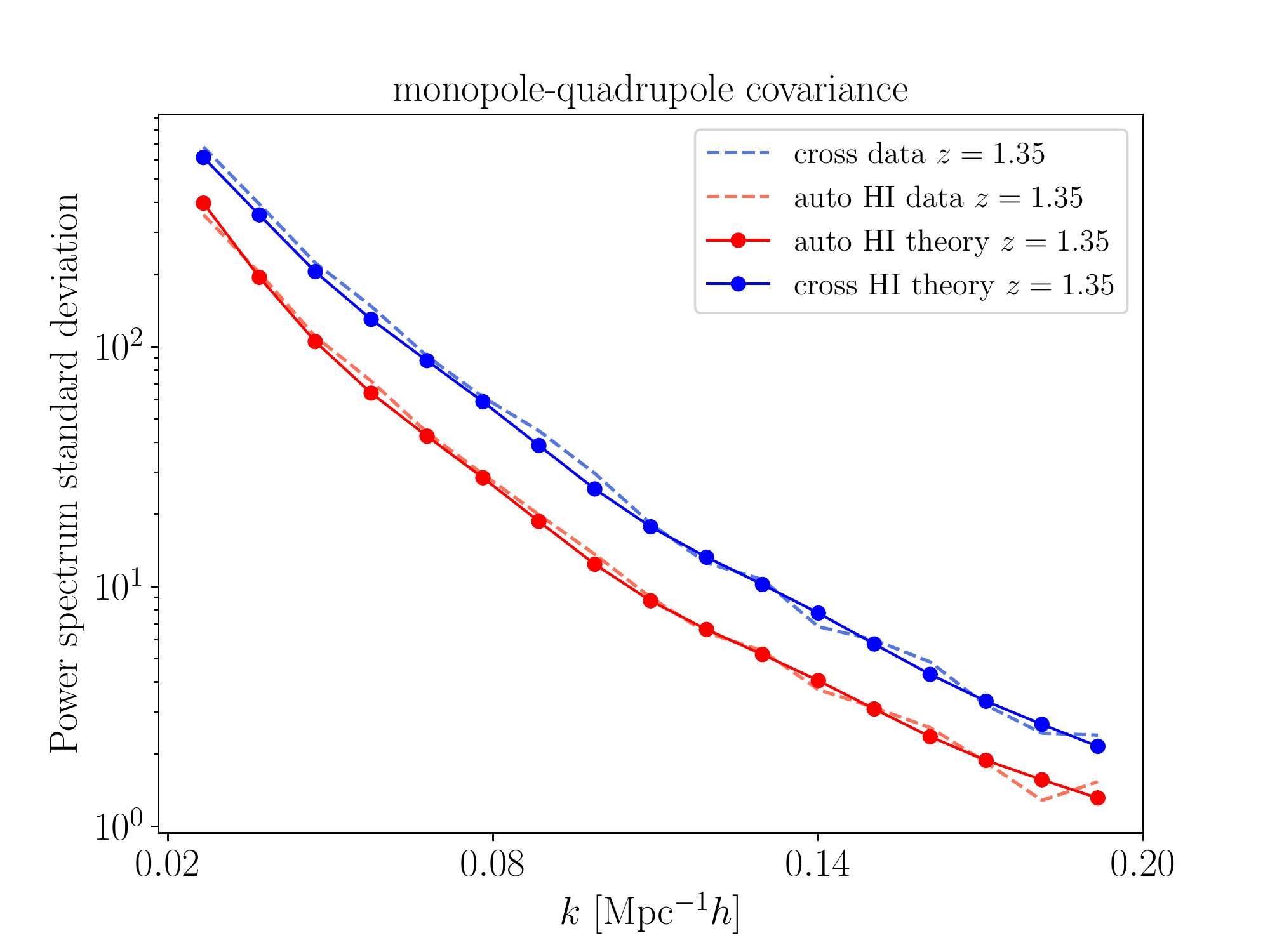}
   \caption{Data variance and theoretical uncertainties at $z=1.35$.}
   \label{fig:poles_gaussian_variance_135}
\end{figure}

The agreement of the data variance with the Gaussian analytic uncertainties can be shown for each component of the covariance matrix. In particular, in \autoref{fig:poles_gaussian_variance_135} we display the theoretical curve (connected filled circles) against the square root of the variance of our $100$-strong data set (dashed). Note that, in order to be shown together, auto- and cross-correlation are normalised to the `thermal' prefactors, thus retaining only the $\mathrm{Mpc^3}\ h^{-3}$ units.

\subsubsection{Signal-to-Noise Ratio}
\begin{figure}
    \centering
    \includegraphics[width=0.9\columnwidth]{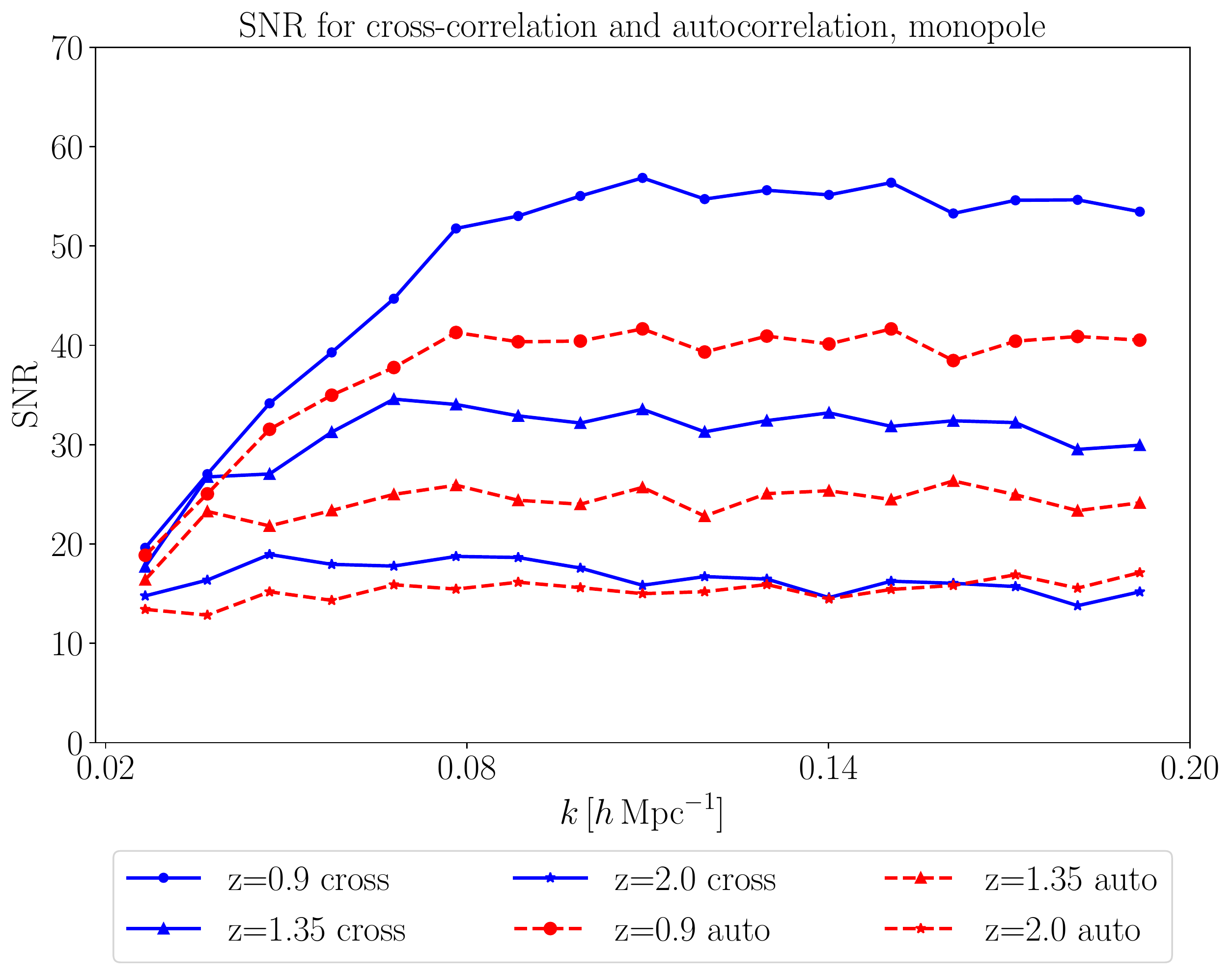}
    \includegraphics[width=0.9\columnwidth]{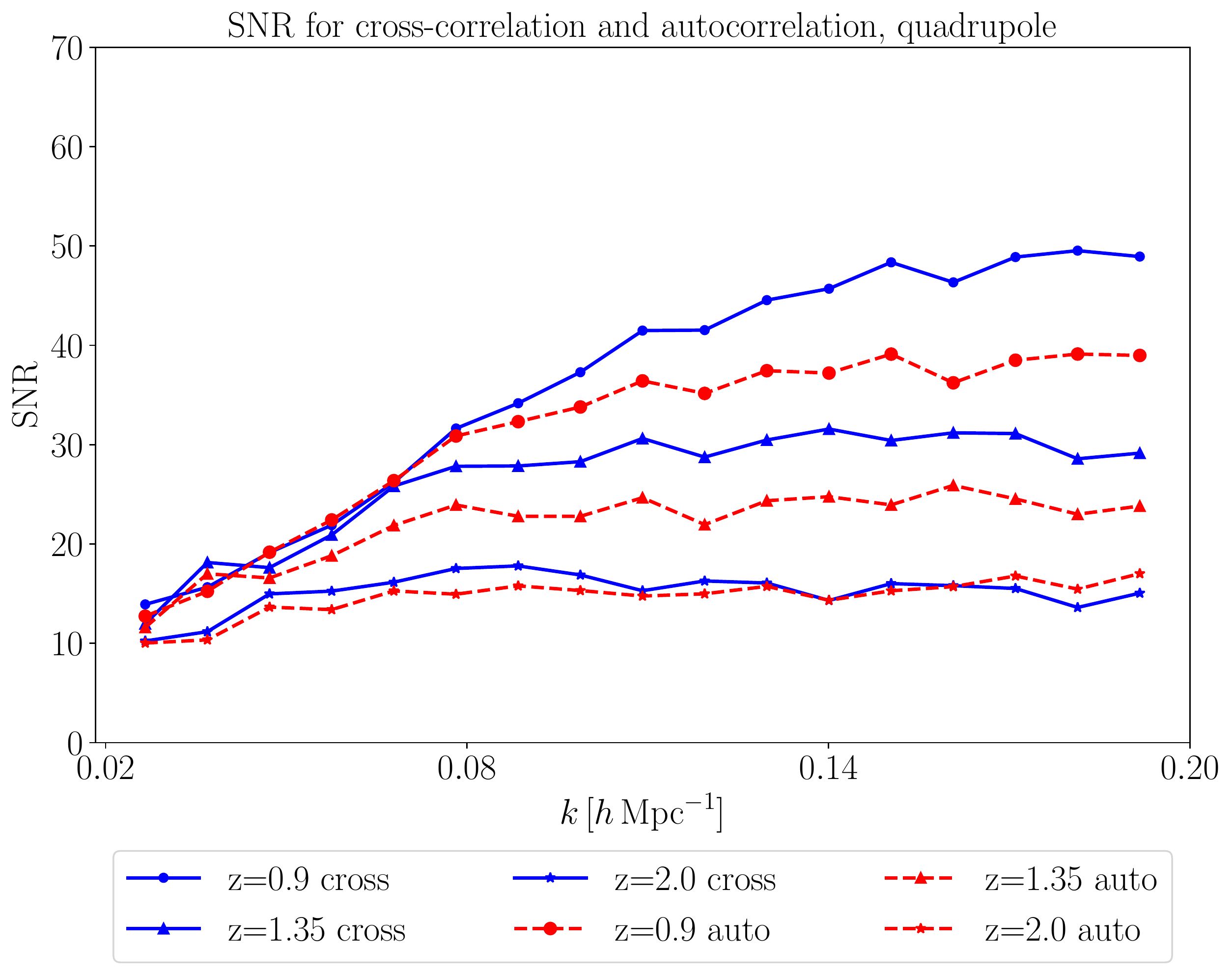}
    \includegraphics[width=0.9\columnwidth]{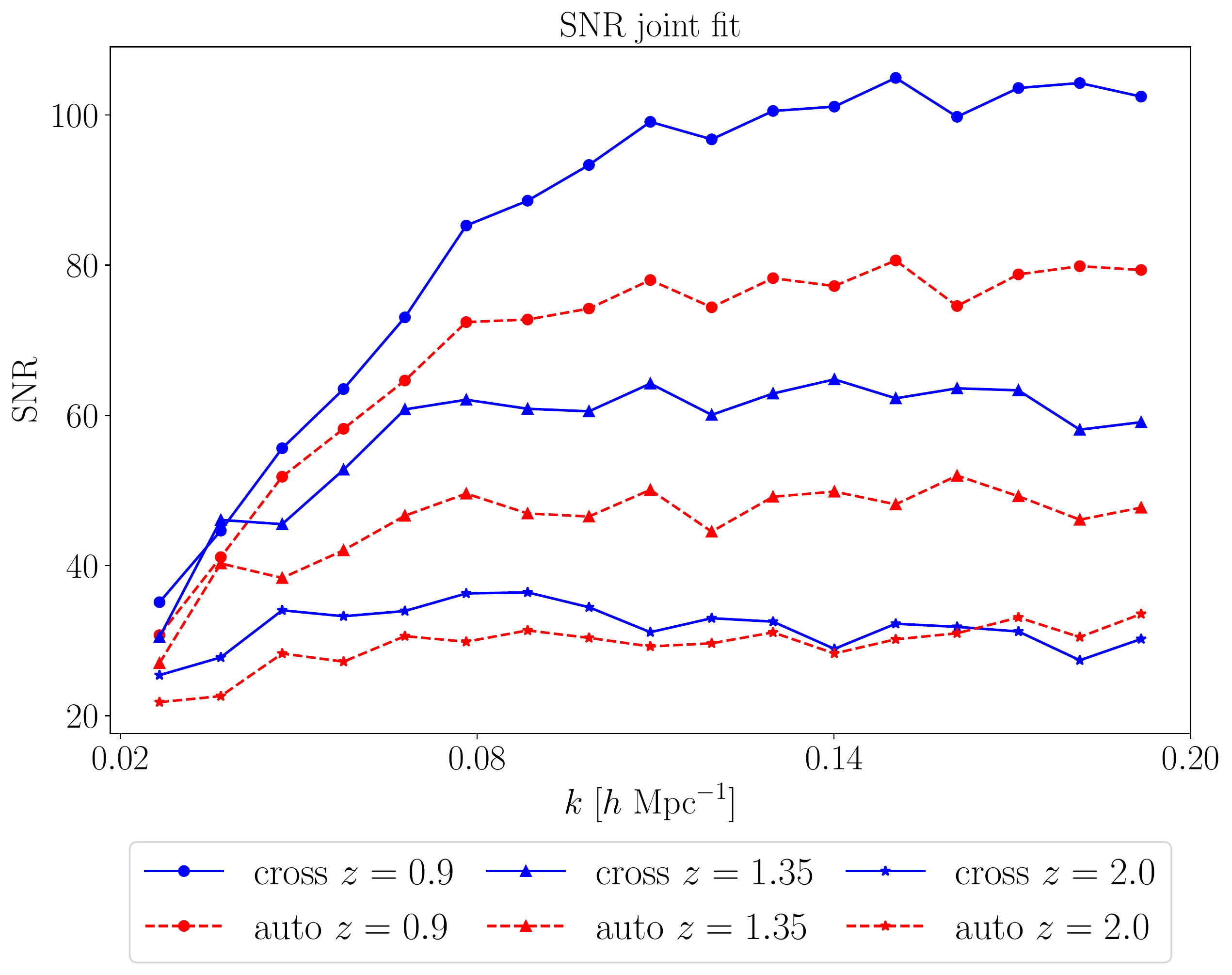}
    \caption{SNR levels for the monopole (left), quadrupole (center) and monopole-quadrupole joint fit (right).} \label{fig:SNR}
\end{figure}

We move now to the monopole and quadrupole SNR, shown in \autoref{fig:SNR}. This quantity appears a particularly useful tool to interpret the behaviours of cross- and auto-correlation described in the previous sections.
The cross-correlation curve lays always above the corresponding auto-correlation SNR, but, while the latter is substantially flat for every $k$, the former exhibits a descending trend in the highest redshift bins and at the highest $k$.
The increasing weight of the shot-noise term, absent in the auto-correlation formula, can explain most part of this trend: increasing $z$, it reduces the distance between auto- and cross-correlation SNR, making positive detection numbers more similar, and bends the latter curve downwards. 

Concerning differences between multipoles, the monopole plateaus are characterised by a higher value and a steeper growth at low $k$ compared with the quadrupole, which reaches the plateau at a slower pace. This can explain why, in spite of the BAO being more visible in the quadrupole, the significance of their detections proves lower: having larger error bars, the $\chi^2$ for the quadrupole null-model improves, reducing the difference between the two templates.
Finally, we can appreciate the overall higher SNR levels reached in the joint fit case: in this case, the quantity is defined as
\begin{equation}
    {\rm SNR}=\sqrt{\bm{\Theta}^{\sf T}\,\mathbfss C^{-1}\, \bm{\Theta}}\;,\label{eq:snr_def}
\end{equation}
$\bm{\Theta}$ being the full data vector i.e.\ a stacking of the $P_{0}$ and $P_{2}$ data vectors, and $\mathbfss C$ the joint monopole-quadrupole covariance matrix.
The three terms of the monopole, quadrupole, and the cross-covariance all positively contribute to the final result.
To conclude, we observe that the SNR scales down with the redshift, a result analogous to what found in \citep[][]{Villaescusa-Navarro:2016kbz} on another power spectrum definition.
Also similarly to that paper, we cannot expect that larger beams would reduce the error bars, for the damping acts more effectively on the power spectrum amplitude, resulting in a disturbance factor for future observations not only in terms of resolution, but also regarding the accuracy of the measurements.

\section{Differences in behaviour between monopole and quadrupole} \label{Appendix_mono_vs_quadru}
\begin{figure*}
    \includegraphics[width=0.7\textwidth]{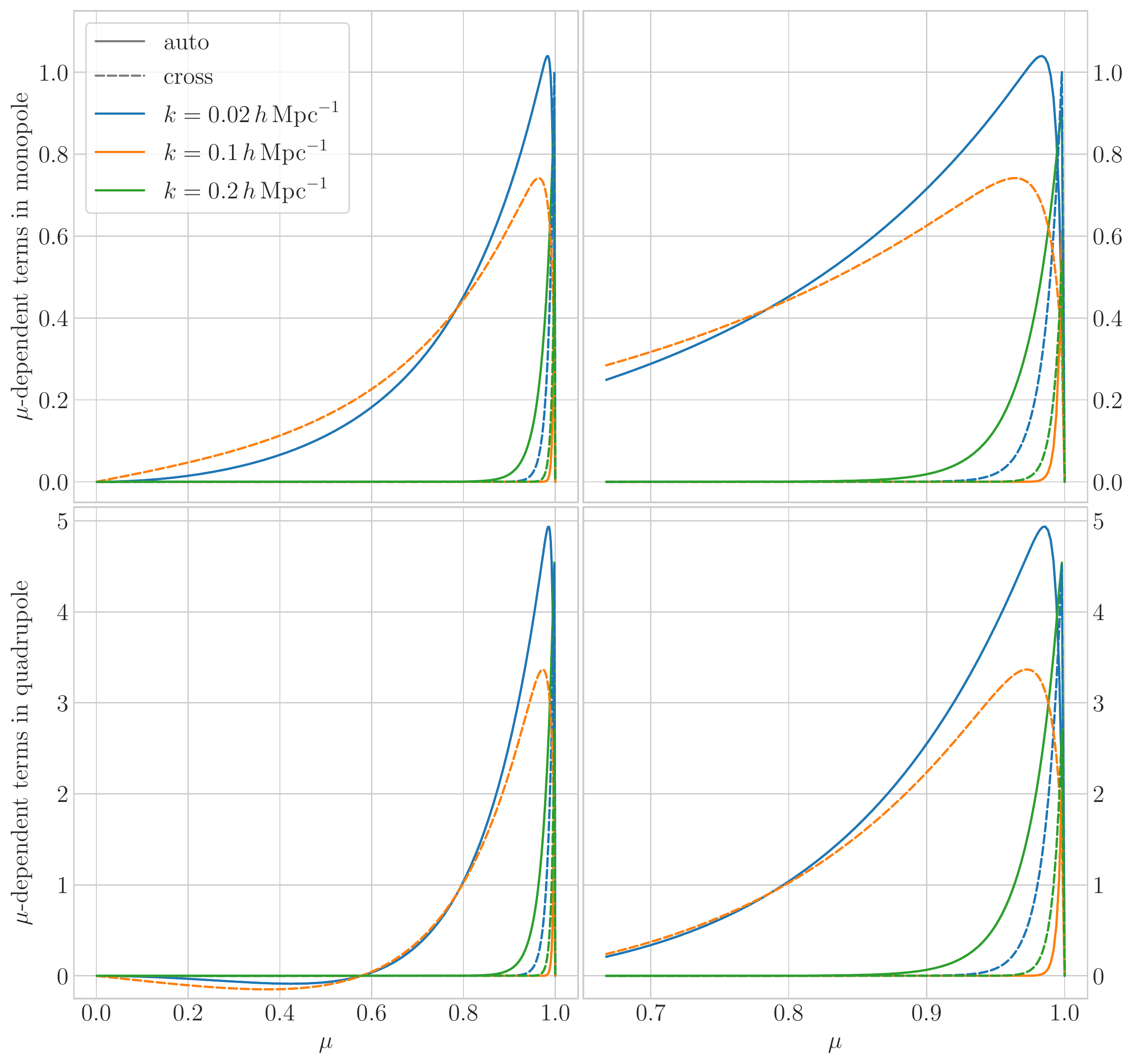}
    \caption{$\mu$-dependent terms for the monopole and quadrupole in either auto and cross-correlation cases, calculated at different scales. Right-hand plot shows the same results zoomed-in on a high-$\mu$ range.}
    \label{fig:mono_quadru_f(mu)}%
\end{figure*}
Together with the modelling differences between auto- and cross-correlation, this work also explores and exploits the different properties of monopole and quadrupole and the influence they have on the BAO amplitudes. 

A useful insight to understand their qualitative, and consequently quantitative, dissimilarities can be obtained by plotting all those $\mu$-dependent terms (RSD, beam factors, compensation windows, $\mathcal{L}_{\ell}$) that appear in the calculation of the power spectrum. To enhance readability, we choose the $z=2.0$ case, displayed in  \autoref{fig:mono_quadru_f(mu)}.
For most modes in the BAO region, and for both monopole and quadrupole, $\mu-$dependent terms assume small or even negative values. By looking at the rightmost plot, where we zoom in the high $|\mu|$ region, i.e.\ along the line of sight, we can better understand how those terms are shaped: their decay towards zero has a similar steepness in both multipoles but the quadrupole starts from higher, and above unity, values than the monopole counterpart, thus enhancing the signal. 
Incidentally, we observe that the relevant contribution to the signal arriving from the line of sight can be connected with the interesting applications of the radial power spectrum outlined in \citep[][]{Villaescusa-Navarro:2016kbz}.

\bsp	
\label{lastpage}
\end{document}